\documentclass[aps,twocolumn,nofootinbib,preprintnumbers,superscriptaddress,eqsecnum]{revtex4}

\pdfoutput=1

\usepackage[
      colorlinks=true,
      linkcolor=blue,
      urlcolor=blue,
      filecolor=black,
      citecolor=red,
      pdfstartview=FitV,
      pdftitle={},
        pdfauthor={Thomas Faulkner, Hong Liu, Mukund Rangamani},
        pdfsubject={},
        pdfkeywords={},
        pdfpagemode=None,
        bookmarksopen=true
      ]{hyperref}

\usepackage[normalem]{ulem}
\usepackage{amsmath}
\usepackage{enumerate}
\usepackage{amsfonts}
\usepackage{color}

\usepackage{graphicx}
\usepackage{bm}

\def\({\left(}
\def\){\right)}
\def\[{\left[}
\def\]{\right]}
\def\<{\langle}
\def\>{\rangle}

\def\CO{{\cal O}}

\newcommand\half{{\ensuremath{\frac{1}{2}}}}
\newcommand\p{\ensuremath{\partial}}

\newcommand\field[1]{{\ensuremath{\mathbb{{#1}}}}}

\newcommand\vev[1]{{\ensuremath{\left\langle{#1}\right\rangle}}}

\newcommand{\RR}{\field{R}}

\newcommand{\be}{\begin{equation}}
\newcommand{\ee}{\end{equation}}
\newcommand{\bea}{\begin{eqnarray}}
\newcommand{\eea}{\end{eqnarray}}
\newcommand{\bwt}{\begin{widetext}}
\newcommand{\ewt}{\end{widetext}}

\newcommand{\bi}{\begin{itemize}}
\newcommand{\ei}{\end{itemize}}
\newcommand{\ben}{\begin{enumerate}}
\newcommand{\een}{\end{enumerate}}
\newcommand{\bca}{\begin{cases}}
\newcommand{\eca}{\end{cases}}
\newcommand{\bln}{\begin{align}}
\newcommand{\eln}{\end{align}}
\newcommand{\bst}{\begin{split}}
\newcommand{\est}{\end{split}}

\newcommand\al{{\alpha}}
\newcommand\ep{\epsilon}
\newcommand\sig{\sigma}

\newcommand\lam{\lambda}
\newcommand\Lam{\Lambda}
\newcommand\om{\omega}

\newcommand\ga{{\ensuremath{{\gamma}}}}
\newcommand\Ga{{\ensuremath{{\Gamma}}}}
\newcommand\de{{\ensuremath{{\delta}}}}
\newcommand\De{{\ensuremath{{\Delta}}}}
\newcommand\vp{\varphi}

\newcommand\ka{{\kappa}}

\newcommand\ov{\over}
\newcommand\ha{{\half}}

\def\le{\left}
\def\ri{\right}

\newcommand\sD{{\ensuremath{{\mathcal D}}}}

\newcommand\sH{{\ensuremath{{\mathcal H}}}}
\newcommand\sL{{\ensuremath{{\mathcal L}}}}

\newcommand\sN{{\ensuremath{{\mathcal N}}}}
\newcommand\sO{{\ensuremath{{\mathcal O}}}}

\newcommand{\vx}{{\vec x}}

\definecolor{rust}{rgb}{0.8,0.2,0.2}
\definecolor{green}{rgb}{0.1,0.8,0.2}

\def\req#1{(\ref{#1})}

\begin{document}

\title{Integrating out geometry: \\
Holographic Wilsonian RG  and the  membrane paradigm}
\preprint{MIT-CTP/4185, NSF-ITP-09xx, DCPT-10/47}

\author{Thomas Faulkner}
\affiliation{Kavli Institute for Theoretical Physics, University of California, Santa Barbara, CA 93106}

\author{Hong Liu}
\affiliation{Center for Theoretical Physics, \\
Massachusetts Institute of Technology,
Cambridge, MA 02139 }
\author{Mukund Rangamani}
\affiliation{Centre for Particle Theory \& Department of Mathematical Sciences, Science Laboratories, South Road, Durham DH1 3LE, United Kingdom }

\begin{abstract}

We formulate a holographic Wilsonian renormalization group flow for strongly coupled systems with a gravity dual, motivated by the need to extract efficiently low energy behavior of such systems. Starting with field theories defined on a cut-off surface in a bulk spacetime, we propose that integrating out high energy modes in the field theory should correspond  to integrating out a part of the bulk geometry. We describe how to carry out this procedure in practice in the classical gravity approximation using examples of scalar and vector fields. 
By integrating out bulk degrees of freedom all the way to a black hole horizon, this formulation  defines a refined version of the black hole membrane paradigm. Furthermore, it also provides a derivation of the semi-holographic description of low energy physics.

\end{abstract}

\today

\maketitle
\tableofcontents

\section{Introduction}
\label{s:intro}

Interacting many-body systems underlie many important physical phenomena. While such systems involve complicated dynamics of a huge multitude of constituents,  one is more often than not  interested in the macroscopic behavior on large distance and  time scales. Fortuitously, in this regime a system generically exhibits universalities which are  independent of specific underlying microscopic dynamics.  This insensitivity to short-distance physics may be understood  using the Wilsonian renormalization group~\cite{Wilson:1973jj,Wegner:1972ih,Wilson:1993dy, Polchinski:1983gv} where microscopic physics is increasingly ``integrated out''  down until the scales of interest. The resulting low energy theory is  only sensitive to a small number of relevant and marginal couplings around some fixed point. 

For systems with a gravity dual, the AdS/CFT correspondence \cite{Maldacena:1997re,Gubser:1998bc,Witten:1998qj} provides a striking geometric picture for the renormalization group flow and the resulting low energy behavior; the radial direction in the bulk can be associated with the energy scale of the boundary theory~\cite{Maldacena:1997re,Susskind:1998dq,Peet:1998wn}, and the radial flow in the bulk geometry can be interpreted as the renormalization group flow of the boundary theory~\cite{Akhmedov:1998vf,Alvarez:1998wr,Girardello:1998pd, Distler:1998gb,Balasubramanian:1999jd,Freedman:1999gp}. This observation has spurred much activity towards a precise formulation of holographic renormalization group (RG), {\it e.g.},~\cite{deBoer:1999xf,deBoer:2000cz}~(see also~\cite{Lewandowski:2002rf,Lewandowski:2004yr}). Recent attempts to derive a holographic duality
directly from field theory concentrate on this very structure \cite{sslee,Lee:2010ub,Douglas:2010rc}.

\begin{figure}[t]
\begin{center}
\includegraphics[scale = 0.54]{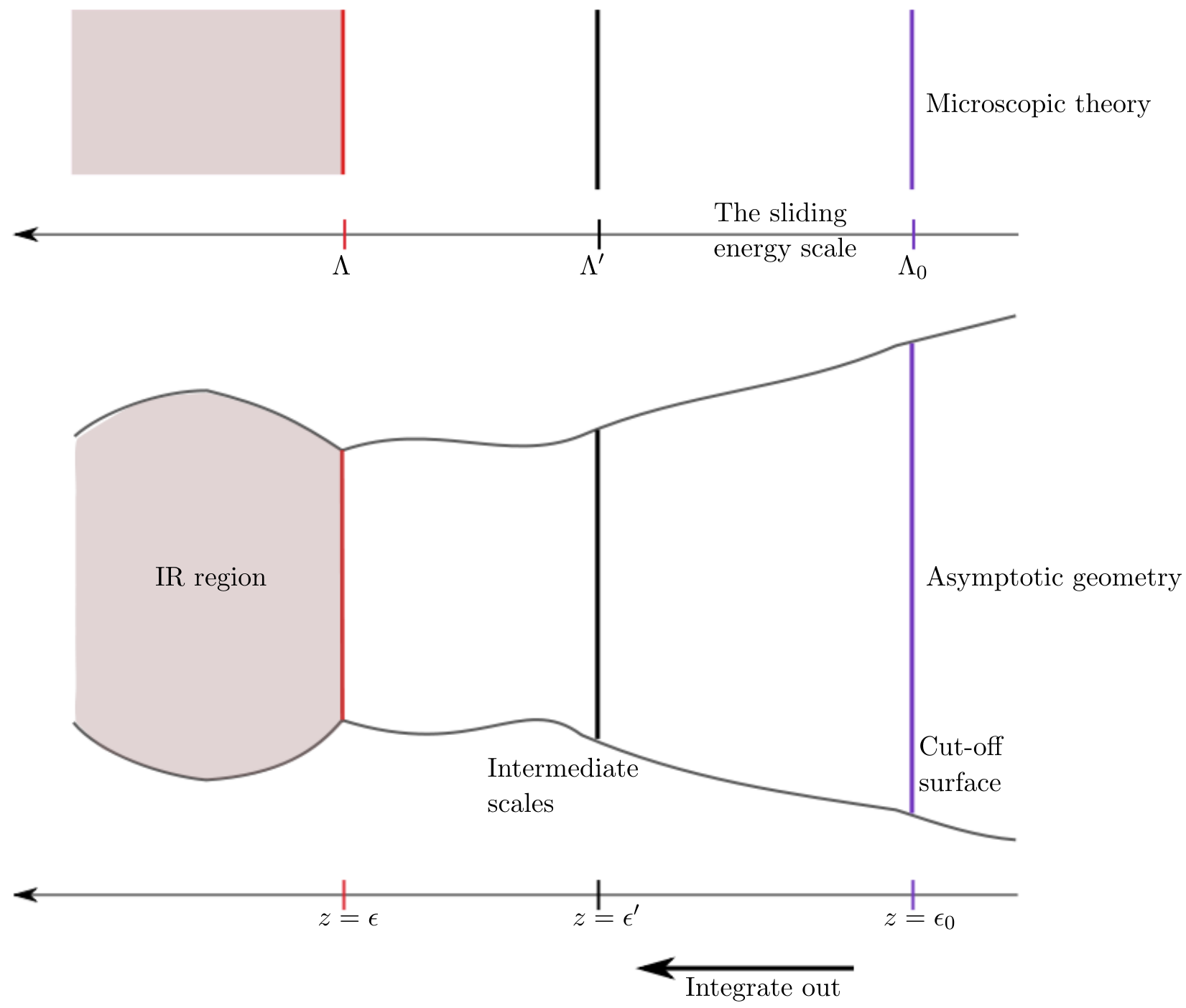}
 \caption{ A schematic description of the effect of integrating out degrees of freedom in the field theory and a corresponding picture of the dual bulk spacetime, with boundary energy scales $\Lam < \Lam' < \Lam_0$ and bulk radial coordinates $\ep > \ep' > \ep_0$.  The boundary of the spacetime lies  in the limit $z \to0$.}
 \label{f:bbh}
\end{center}
\end{figure}

In this paper we propose a formulation of holographic renormalization group flow, motivated by 
the Wilsonian approach of integrating out short-distance degrees of freedom.  The basic idea is as follows. 
We identify the boundary theory defined with a cut-off scale $\Lam_0$ with the bulk theory defined in the spacetime region $z > \ep_0$ for some $\ep_0$ (see Fig.~\ref{f:bbh}).  Integrating out degrees of freedom in the boundary theory from $\Lam_0$ to some lower scale $\Lam'$ is then identified with integrating out the bulk degrees of freedom between $z = \ep_0$ and some $z= \ep' > \ep_0$.\footnote{Note that, as will be discussed in the main text, there is a caveat to this statement: there could exist gapless degrees of freedom in the region of the bulk which is  being integrated out. One should isolate and retain these modes in the  low energy effective theory.}  Integrating out the bulk degrees of freedom in the region $\ep_0 < z <  \ep'$ results in a boundary action $S_B (z=\ep')$ at $z=\ep'$ hypersurface.  $S_B$ provides boundary conditions for bulk modes in the region $z > \ep'$ and can be considered as specifying a ``boundary state'' for the bulk theory in that region.  We propose that this effective action $S_B$ can be identified with the Wilsonian effective action of the boundary theory at the scale $\Lam'$, with couplings in $S_B$ identified with those for single-trace and induced multiple-trace operators in the boundary theory.  Requiring that physical observables be independent of the choice of the cut-off scale $\ep'$ then determines a flow equation for the Wilsonian action $S_B$ and associated  couplings.  
We will restrict our discussion to the classical gravity regime and hence our equations should be viewed as the large $N$ limit of the full flow equations; a more general discussion of this functional equation appears in~\cite{Heemskerk:2010hk}. To illustrate the general idea we will use scalars and vector fields propagating in a fixed spacetime as examples, leaving the analysis of gravitational degrees of freedom for future investigations.  

 The approach proposed here differs from  that in previous literature, {\it e.g.},~\cite{deBoer:1999xf}, where the flow equation was defined for the full classical bulk action in the entire region $z>\ep'$. As a result the flow necessarily involves knowledge of geometry and dynamics in the region interior to the cut-off surface $z=\ep'$,  and thus cannot be Wilsonian. In the approach developed in this paper,  the flow equation is for the boundary effective action $S_B$ which only involves dynamics of the portion of the geometry which is being integrated out,  $z < \ep'$, and knows nothing  about the interior region. 

The holographic Wilsonian RG developed here greatly simplifies the characterization of low energy behavior of a boundary system; as indicated in Fig.~\ref{f:bbh}, one can push the cut-off surface into some relevant infrared (IR) region of the bulk geometry and then forget the rest of the spacetime (see also~\cite{Bredberg:2010ky,Nickel:2010pr} for recent works emphasizing similar ideas). For example, consider a boundary theory at a non-zero temperature or chemical potential, which is described in the bulk by a (charged) black hole. By implementing the process of integrating out the bulk spacetime, one can push the cut-off surface all the way to the near-horizon portion of the black hole geometry, say onto the stretched horizon. The low energy dynamics of the boundary system, including the hydrodynamical regime, should be captured by the dynamics of the stretched horizon coupled to the effective action $S_B$  obtained from integrating out the rest of the geometry. This defines a more refined version 
of the so-called ``membrane paradigm'' \cite{Damour:1978cg,Thorne:1986iy} in classical black hole physics, where it was observed that the stretched horizon behaves as a fictitious fluid. 
Note that in previous studies of the fluid/gravity correspondence (see~\cite{Son:2007vk, Rangamani:2009xk} for reviews) it has been clear ({\it e.g.},~\cite{Bhattacharyya:2008jc, Iqbal:2008by}) that the dynamics of the stretched horizon alone is {\em not} enough to capture the boundary theory hydrodynamics and the rest of the spacetime is needed. In our current approach the effect of the other part of the spacetime is now captured by the effective 
action $S_B$ on the stretched horizon. We illustrate this by recovering the diffusion behavior of 
a conserved current at finite temperature, leaving the study of full-fledged hydrodynamics for the future. 

In a previous attempt to make connection to the membrane paradigm in AdS/CFT~\cite{Iqbal:2008by}, a set of flow equations were derived to relate boundary response functions to those on the stretched horizon. We show that these flow equations precisely correspond to those for double-trace couplings in the present Wilsonian formulation.  

For an extremal charged black hole, the horizon becomes degenerate and the near-horizon region opens up to an AdS$_2$ throat with infinite proper distance (see Fig.~\ref{f:ads2gap} in Sec.~\ref{s:semiholo}). As a result one anticipates that in the IR the system is described by a $(0\!+\!1)$-dimensional  CFT$_1$~\cite{Faulkner:2009wj}. In this case the natural place to put the cut-off surface is the boundary of AdS$_2$ and the effective action $S_B$ at the cut-off surface can be interpreted in the boundary theory as multiple-trace deformations of the CFT$_1$ arising from integrating out UV degrees of freedom. In particular, if there are gapless modes in the part of the geometry that has been integrated-out (i.e., in the region outside the AdS$_2$ throat) one should include such modes in the low energy theory (i.e., in $S_B$) resulting a picture of such modes\footnote{In studies of holographic Fermi surfaces referred above, the modes in question are the free fermions around the Fermi surface.}  coupling to a strongly interacting CFT$_1$. 
This is exactly what happens for holographic Fermi surfaces~\cite{Liu:2009dm,Cubrovic:2009ye,Faulkner:2009wj,Faulkner:2010da} where the present  RG perspective allows us to give a derivation of the semi-holographic picture~\cite{Faulkner:2010tq}.  This discussion extends to other geometries with a nontrivial IR region, for example,  
the ground states for holographic superconductors in AdS$_4$ where the IR region can be 
a Lifshitz geometry or another AdS$_4$~\cite{Gubser:2009cg,Horowitz:2009ij,Gauntlett:2009bh}.

The outline of the paper is as follows:  in Sec.~\ref{s:holwilson} we lay down the general formulation of holographic Wilsonian RG flow using a scalar field and discuss some simple examples. In Sec.~\ref{s:semiholo} we consider the specific example of an extremal black hole, which we use to illustrate the possibility of appearance of gapless modes in the UV region and derive the semi-holographic picture. Sec.~\ref{s:vectors} is devoted to vector fields, where we additionally need to account for issues associated with gauge invariance.  We end in Sec.~\ref{s:conclude} with a discussion and open questions. In Appendix~\ref{app:a} we give an explicit discussion how to relate quadratic terms in the effective action $S_B$ to double-trace deformations in the corresponding CFT. 

{\em Note added:} while this paper is being finalized, we received~\cite{Heemskerk:2010hk}, which overlaps with our discussion in Sec.~\ref{s:holwilson} and Sec.~\ref{s:vectors}.

\section{A formulation of the holographic Wilsonian flow}
\label{s:holwilson}

Consider a field theory defined by a path integral below some UV cut-off $\Lam$ 
\be 
Z = \int_{ \Lam}  D \Phi \, \exp \le[i \,I_{eff}  [\Phi, \Lam]  \ri]
\ee
where $\Phi$ denotes collectively all the fields involved. We will denote local gauge-invariant operators  by $\CO$. 
The effective action $I_{eff} [\Phi, \Lam]$ at the scale $\Lam$  can be written as 
\be \label{nis}
I_{eff} [\Phi, \Lam] = I_0 [\Phi] + I_{UV} [\Phi, \Lam]
\ee
where $I_0[\Phi]$ is the original (microscopic) action and $I_{UV}$ arises from integrating out degrees of freedom above the cut-off scale $\Lam$. In order for $Z$ to be independent of the cut-off $\Lam$,
$I_{UV}[\Lam, \Phi]$ should satisfy a renormalization group flow equation.  When  $I_{UV}$ is expanded  in terms of a complete set of local (gauge invariant) operators, this flow equation then gives the $\beta$-functions for the complete set of couplings.  

Here we are interested in a boundary theory with a gravity dual, say $\sN=4$ super-Yang-Mills theory with a gauge group $SU (N)$, where $I_{UV} [\Phi,\Lam]$ generically contains single and {\em multiple}-trace gauge invariant operators (see {\it e.g.},~\cite{Li:2000ec, Akhmedov:2002gq,Pomoni:2008de, Akhmedov:2010sw,Vecchi:2010jz}). In the $N \to \infty$ limit one expects the flow equation for $I_{UV}$ should dramatically simplify given factorizations of correlation functions in such a limit. In this section we propose a counter-part for $I_{UV}$  in the large $N$ limit in the holographic gravity dual.  We use a bulk scalar field $\phi$ which is dual to a scalar boundary operator $\sO$ for illustration. Generalizations to multiple fields are self-evident and  generalizations to  vector fields will be discussed in Sec.~\ref{s:vectors}.

\subsection{Holographic flow equations}

We work with a static $d+1$ dimensional bulk geometry which is rotationally and translationally invariant along boundary directions, whose metric can be written in a form
\be \label{gmet}
ds^2 = g_{MN} dx^M dx^N \equiv  - g_{tt} dt^2 + g_{ii} d\vx^2 + g_{zz} dz^2 \ 
\ee
with $g_{MN}$ depending on $z$ only and $x^M = (z, x^\mu) = (z, t, x_i)$.  
While our general discussion should be applicable to various asymptotics of the geometry, for definiteness we will concentrate on asymptotically AdS spacetimes, which is attained in the limit $ z \to 0$.

Now consider a scalar field $\phi$ with an action 
\be \label{bact}
S =  \int_{z > \ep} d^{d+1} x \,\sqrt{-g} \,\sL (\phi, \p_M \phi)  + S_B [\phi, \ep]
\ee
where $\sL$ is the bulk Lagrangian and $S_B$ is a boundary action defined at the surface $z= \ep$.  $S_B [\phi, \ep]$ defines the boundary conditions for the bulk field $\phi$ at $z = \ep$. In fact one can view it as  specifying a ``boundary state'' for the bulk theory in the region $z > \ep$. It can be interpreted as coming from integrating out degrees of freedom for $\phi$ for $z < \ep$. 

In all known examples of metric~\eqref{gmet} with a boundary theory dual, $g_{tt}$ is a monotonically decreasing function of $z$, which implies that intervals of  boundary time $\Delta t$ are increasingly red-shifted compared with the local proper time $\Delta \tau \approx \sqrt{g_{tt}} \, \Delta t$ as $z$ is increased (i.e., going deeper into the interior). Thus lower energy processes in the boundary theory are more and more associated with bulk physics in the deeper interior. This is the standard IR/UV connection~\cite{Susskind:1998dq} and gives a natural interpretation of 
the $z$ direction as the renormalization group scale of the boundary theory.\footnote{While it is clear that $z$ should be inversely related to the boundary cut-off scale $\Lam$, for a generic metric their relation could be complicated. For pure AdS (or more generally near the AdS boundary), one has $g_{tt} \propto z^{-2}$ and  $z \propto {1 \ov \Lam}$.} 

We propose that the boundary term $S_B [\phi, \ep]$ be interpreted as dual to the boundary theory effective action $I_{UV}$ in equation~\eqref{nis}.  Their relation is particularly simple in the so-called `alternative quantization'~\cite{Breitenlohner:1982jf},  where  the boundary value of the scalar field  $\phi$ is identified (in the absence of $S_B$) as the expectation value of a dual  operator operator $\sO$~\cite{Klebanov:1999tb}. Thus $S_B [\phi, \ep]$ should directly translate into the boundary theory as $I_{UV}$ (up to some renormalization). 
 In particular, if we expand $S_B$ in power series of $\phi$, then the linear term will correspond to the dual operator $\sO$, $\phi^2$ terms will correspond to double-trace operator $\sO^2$, and $\phi^n$ to multiple-trace operators $\sO^n$, etc.. In contrast for the standard quantization, where the boundary value of $\phi$ is interpreted as the source, $I_{UV} [\sO]$ should be identified with the Legendre transform of $S_B$ (again up to some renormalization)~\cite{Klebanov:1999tb}. Note that far away from a fixed point, the identification of $S_B$ with $I_{UV}$ is likely not unambiguous and will depend on the renormalization scheme etc.. In particular, there should be no distinction between the standard and alternative quantizations;  the above two descriptions should be equivalent, although depending on specific situations, one may be more convenient than the other. Near a fixed point, the above discussion can nevertheless be made precise, see equations~\eqref{in1}--\eqref{in2} in Appendix~\ref{app:a} for explicit expressions, where we also illustrate this explicitly using a simple example of double- trace deformations. 
  
There is an important caveat in the above identification of $S_B$ with $I_{UV}$, 
as one cannot really make a precise identification of integrating out the bulk degrees of freedom for $z < \ep$ with integrating out boundary degrees of freedom above some cut-off scale $\Lam$. After all, for any boundary physical process (no matter what energy), all regions in the bulk contribute. In particular, $I_{UV}$, coming from integrating out high energy degrees of freedom, has a well-defined expansion in terms of local operators. But this is not necessarily the case for $S_B$. Various examples are known in which gapless (or close to gapless) modes exist in the UV region, including modes near a holographic Fermi surface~\cite{Liu:2009dm,Cubrovic:2009ye,Faulkner:2009wj,Faulkner:2010da}, ``Goldstone modes'' in a symmetry breaking phase (see {\it e.g.},~\cite{Iqbal:2010eh,Nickel:2010pr}), order parameters close to a phase transition~\cite{Faulkner:2010gj}. Integrating out these modes may induce non-local terms in $S_B$ and its Legendre transform. In order to have a proper description of IR dynamics, one should isolate these gapless modes from $S_B$ and treat them separately. We will discuss explicit examples of this in detail in Sec.~\ref{s:semiholo} and Sec.~\ref{s:vectors}. In the absence of such gapless modes (or after subtracting them from $S_B$), we expect it should be possible to identify $S_B$ with $I_{UV}$ in some specific cut-off scheme of the boundary theory, although the precise specification of such a scheme and  a precise relation between $\Lam$ and $\ep$ will most likely be difficult to obtain in general.\footnote{Given that there are also high energy modes in the region $z > \ep$ which are unintegrated, the corresponding field theory description should involve some kind of soft cut-off.}

We will now derive a flow equation for $S_B$ by requiring that physical observables are independent of $\ep$ as it is varied. To be specific we restrict our discussion to the classical gravity limit which corresponds to the large $N$ (planar) limit of the boundary field theory. For definiteness we take the Lagrangian in~\eqref{bact} to be 
\be \label{iwpw}
\sL = - \ha (\p \phi)^2 - V (\phi)  \ .
\ee
Varying the action we find that the equation of motion
\be \label{eom}
{1 \ov \sqrt{-g}} \p_M \le(\sqrt{-g} g^{MN} \p_N \phi \ri) - {\p V \ov \p \phi} = 0
\ee
with boundary condition (evaluated at $z = \ep$)
\be \label{bde}
\Pi =  {\de S_B \ov \de \phi}, \qquad  \Pi \equiv  -\sqrt{-g}\,g^{zz}\, \p_z \phi
\ee
where $\Pi$ is the canonical momentum along the radial direction. In the the large $N$ limit of the field theory we are interested in the value of the on-shell action $S_{cl}$ evaluated on a solution to~\eqref{eom}--\eqref{bde}. 

Since the choice of our cut-off surface $z = \ep$ is arbitrary, the physical requirement of demanding that the on-shell action $S_{cl}$ evaluated on a solution (and the solution itself) does not change, imposes a flow equation for the boundary action $S_B$, i.e.,
\be
0 =  - \!\int_{z=\ep} \!\! d^d x \, \sqrt{-g} \, \sL   + \p_\ep S_B [\phi, \ep] 
+ \int_{z = \ep}  d^d x \, {\de S_B  \ov \de \phi (x)} \p_z \phi (x)  .
\ee
Using~\eqref{bde} the above equation can be written as
\be \label{ff1}
\p_\ep S_B [\phi, \ep]   =  
- \int_{z=\ep}  d^d x \, \le( \Pi \p_z \phi   -\sqrt{-g}\,  \sL \ri) = - \int d^d x \, \sH 
\ee
where $\sH$ is the Hamiltonian density for evolution in the $z$ direction. The above equation is intuitively 
clear; the flow is generated by the Hamiltonian and is governed by the Hamilton-Jacobi equation. 
Writing out $\sH$ explicitly and using~\eqref{bde} we can also write the flow equation as
 \bea
 \sqrt{g^{zz}} \p_\ep S_B [\phi, \ep]   &=&  - \int_{z=\ep}  d^d x \, \sqrt{-\ga} \, \le({1 \ov 2 \ga} \le({\de S_B \ov \de \phi}\ri)^2  \ri. \cr
&&  \le. +  \ha g^{\mu \nu} \p_\mu \phi \p_\nu \phi  + V(\phi) \ri)
 \label{rfed}
\eea
where
\be \label{defge}
\ga \equiv \det g_{\mu \nu} = g g^{zz} \ .
\ee
Note that we should treat~\eqref{rfed} as a functional equation and in particular should not impose the equation of motion~\eqref{eom} when evaluating it. When $S_B$ is expanded in power series of $\phi$, the equation specifies how the coefficients of the expansion flow with $\ep$. It is important to emphasize that equation~\eqref{rfed} is a flow equation for the boundary action $S_B$ rather than the full classical action $S_{cl}$ as discussed in previous literature, {\it e.g.},~\cite{deBoer:1999xf,deBoer:2000cz}. Similar flow equations were considered in~\cite{Lewandowski:2002rf,Lewandowski:2004yr} for the effective action on the Planck brane in the Randall-Sundrum compactification.  An equation similar to~\eqref{rfed} was also considered slightly earlier in~\cite{Papadimitriou:2010as} in the context of holographic renormalization, but a Wilsonian interpretation was not given. 

\subsection{Extracting low energy behavior: where to put the cut-off surface}

The holographic Wilsonian RG approach outlined here can be used to simplify the task of extracting low energy behavior of the dual theory. Recall that in the standard formulation of AdS/CFT duality, the generating functional of correlation functions in the  boundary theory is given by~\cite{Gubser:1998bc,Witten:1998qj} 
\be \label{stanD1}
e^{I [J]} \equiv \vev{e^{\int J \sO} } =  \lim_{\ep \to 0} e^{S_{0} [\phi_c, z \geq \ep ] + S_{ct} [\phi_c, z=\ep]}  . 
\ee 
Here $S_0$ is the bulk action given in~\eqref{iwpw} and  $S_{ct}$ is a counter-term action (required to ensure a well defined variational principle). The field $\phi_c$ is a classical solution satisfying  appropriate boundary conditions. These take the form of a regularity (or in-falling) condition in the interior of the spacetime and  an asymptotic  boundary condition specified by the source $J$ (which generically is either Dirichlet or Neumann). From a field theory perspective one can imagine the above as prescribing data at the fixed point.

Given this set-up one can solve the  flow equation~\eqref{rfed} to determine $S_B$ at some scale $\ep$. One integrates the field in the bulk starting with the initial data specified  at $ \ep =0$ for the field $\phi_c$ as in~\eqref{stanD1}. This process ensures that the generating functional is given by
\be \label{oepr1}
e^{I [J]} =   e^{S_{0} [\phi_c, z \geq \ep] + S_{B} [\phi_c, z=\ep]} \ 
\ee
for any $\ep$. The key difference of course is that  now $\phi_c$ is found by satisfying the boundary condition at $z = \ep$ specified by $S_B$ (clearly the flow equation of $S_B$ also ensures that $ \phi_c$ obtained this way is the same as that given in~\eqref{stanD1}). 

If one is interested in obtaining full correlation functions for arbitrary momentum and frequency, equation~\eqref{oepr1} by itself does not offer any simplifications compared to~\eqref{stanD1} as solving for $S_B [\ep]$ is equivalent to solving the classical equation of motion in the integrated out region. This is not much of a surprise, the same statement would be true in field theory; integrating out momentum shells does not simplify computations should one be interested in extracting physics at arbitrary scales.

However, equation~\eqref{oepr1} does offer much simplification in extracting low frequency $\om$ (and/or small momentum $k$) behavior of correlation functions.  For such purpose we can expand $S_B$ analytically in small $\om$ and/or $k$ analogous to what one would do with the Wilsonian effective action in field theory. The leading order expression in such an expansion is often not difficult to obtain, as we will see in various examples in the following. 

More interestingly, the expansion in $\omega$ (and/or $k$) also determines where we should 
put the cut-off surface $z=\ep$; it should be put at the boundary of some IR region where analytic expansion in $\om$ or $k$ breaks down. Such a breakdown signals the presence of new light degrees of freedom that must be retained in the low energy dynamics and not be integrated out. 
For example, in the geometry of a black hole with a non-degenerate horizon, we can put the cut-off surface just outside the horizon, while for an extremal black hole, it should be put at the boundary of the near horizon AdS$_2$ region, as the analytic expansion in $\om$ breaks down in the AdS$_2$ region. Effectively one is isolating a region of the geometry which has dominant contribution to low energy physics of the field theory; as a consequence we are able to formulate a refined version of the membrane paradigm and derive the semi-holographic models of low energy effective field theories.

\subsection{Holographic Wilsonian flow for a free scalar}
\label{s:howfs}

We now illustrate the flow equation~\eqref{rfed} more explicitly by considering a free bulk theory
with $V (\phi) = \ha m^2 \phi^2$ and expanding $S_B$ in momentum space as:
\bea 
S_B [\ep, \phi] &=& \Lam (\ep) + \int {d^d k \ov (2 \pi)^d } \, \sqrt{-\ga}\,  J(k, \ep) \phi (-k) \cr
 && \hspace{-4mm} - \ha \int  {d^d k \ov (2 \pi)^d} \,  \sqrt{-\ga}\, f(k,\ep) \,\phi (k) \phi (-k)  \ 
\label{wexp}
\eea
where as mentioned earlier and also elaborated more in Appendix~\ref{app:a}, $f(k)$ is related to couplings for double-trace operators made from $\sO$. Here and below we use the following notation:  
\bea 
k_\mu = (-\om, k_i), \qquad d^d k = d\om \,d ^{d-1} k_i , \qquad \cr
k^2 \equiv \sum_i k_i^2, \qquad k^\mu k_\mu = - g^{tt} \om^2 + g^{ii} k^2 \ .
\eea

 Plugging~\eqref{wexp} into~\eqref{rfed} we then find that
\be \label{oi1}
{\cal D}_\ep  \Lam  =  \ha \int {d^d k \ov (2 \pi)^d} \, J(k,\ep) J(-k,\ep) ,
\ee
\be \label{lo1}
{\cal D}_\ep  \le(\sqrt{-\ga} \;J(k, \ep) \ri)  = - J(k,\ep)\; f(k,\ep) ,
\ee
\be \label{flo}
{\cal D}_\ep  \le( \sqrt{-\ga}\; f (\ep, k)\ri)  = - f^2(k,\ep) + k^\mu k_\mu + m^2 
\ee
where
\be \label{defde}
{\cal D}_\ep = {1 \ov \sqrt{-\ga}} \sqrt{g^{zz}} \,\p_\ep = \frac{1}{\sqrt{-g}}\, \p_\ep \ .
\ee

By specifying the initial condition $\Lam_0 \equiv \Lam(\ep_0), J_0 \equiv J(\ep_0)$ and $f_0 \equiv f(\ep_0)$ on some initial surface $z = \ep_0$, equations~\eqref{oi1}--\eqref{flo} can then  be used to determine these quantities at some other surface at $z = \ep > \ep_0$.  Since we are working with a classical theory, we expect that equations~\eqref{lo1}--\eqref{flo} should be related to the classical equations of motion~\eqref{eom}.\footnote{Once $J$ is known,~\eqref{oi1} can be integrated directly.} Indeed writing equation~\eqref{eom} in the first order form 
\be \label{eom1}
\p_z \phi = - {g_{zz} \ov  \sqrt{-g}} \Pi
\ee 
\be \label{eom2}
\p_z \Pi =-  \sqrt{-g}  \le( k_\mu k^\mu +  m^2 \ri) \phi
\ee
it can be readily checked that given a solution $\phi_s$ and $\Pi_s$ satisfying~\eqref{eom1}--\eqref{eom2} we obtain a solution of~\eqref{lo1}--\eqref{flo} by setting
\be \label{pepw}
f = - {\Pi_s \ov \sqrt{-\ga} \phi_s} , \qquad J = {1 \ov \sqrt{-\ga} \phi_s} \ . 
\ee  
Note that a version of equation~\eqref{flo} was derived earlier in~\cite{Iqbal:2008by} in considering the black hole membrane paradigm in AdS/CFT, where its interpretation in terms of RG flow was only speculated upon. Using the formalism of the holographic Wilsonian flow we are now able to give a precise interpretation for it as the $\beta$-function equation for double-trace couplings.

Now consider a basis of independent solutions $\phi_1 (z), \phi_2(z)$ to equations~\eqref{eom1}--\eqref{eom2} with corresponding canonical momenta given by $\pi_1 (z), \pi_2 (z)$.\footnote{As the equations are real, we can take $\phi_{1,2}$ to be real.}  Using~\eqref{pepw}, we can then find an explicit expression for  $J (\ep)$ and $f (\ep)$ in terms of the initial conditions $f_0, J_0$ at $z = \ep_0$ 
\be
\begin{split} \label{sl2T}
\sqrt{-\ga} J(\ep) &= {\sqrt{-\ga_0} J_0 \ov u (\sqrt{-\ga_0} f_0) + v} , \cr 
\sqrt{-\ga} f (\ep)  & = {r (\sqrt{-\ga_0} f_0)  + s \ov u (\sqrt{-\ga_0} f_0)  +v}
\end{split}
\ee
where $\ga_0 \equiv \ga (z=\ep_0)$ and  
\be \label{tras}
\left( \begin{matrix} r&s\\ u&v \end{matrix} \ri) = M (\ep) M^{-1} (\ep_0) 
\ee
with $M$ defined by
\be
M (z) \equiv  \left( \begin{matrix} -\pi_1 (z) & -\pi_2 (z) \\ \phi_1 (z) &\phi_2 (z) \end{matrix} \ri) \ .
\ee
The matrix $\left( \begin{smallmatrix} r&s\\ u&v \end{smallmatrix} \ri)$
has determinant $1$ (thus belongs to $SL(2,{\mathbb R})$), since
\be \label{rons}
\det M (z) = \phi_1 (z) \pi_2 (z) - \phi_2 (z) \pi_1 (z) \equiv W
\ee 
is the Wronskian for $\phi_1, \phi_2$ and is $z$-independent. 

An alternative way to write~\eqref{sl2T}, which is sometimes more convenient, is as follows. 
The classical solution $\phi_s$ to~\eqref{eom1}--\eqref{eom2} can be expanded in terms of the basis $\phi_{1,2}$ as   
\be  \label{oepr}
\phi_s = \al \,\phi_1 + \beta \,\phi_2 \ .
\ee
Specifying $(\al, \beta)$ is equivalent to specifying $(f_0, J_0)$ and the advantage of using $(\al,\beta)$ is that they are integration constants which are invariant under the flow. More explicitly, for any $z$, we have 
\be \label{abc}
\left( \begin{matrix} \al \\ \beta \end{matrix} \ri) = M^{-1} (z) {1 \ov \sqrt{-\ga} J }  \left( \begin{matrix}\sqrt{-\ga} f\\ 1 \end{matrix} \ri) \ .
\ee
 In fact, instead of $(\al,\beta)$, it is slightly more convenient to consider $(\chi \equiv {\beta \ov \al}, \al)$, for which~\eqref{abc} becomes 
 \be  \label{flowinv}
\chi \equiv  {\beta \ov \al} = - {\phi_1 \sqrt{-\ga} f + \pi_1 \ov \phi_2 \sqrt{-\ga} f + \pi_2 }, \quad
\al = {\phi_2 \sqrt{-\ga} f + \pi_2 \ov W \sqrt{-\ga} J}
\ee
where $W$ was introduced in~\eqref{rons}. Note that $\chi$ only depends on $f$. 
Inverting~\eqref{flowinv}, we then have 
 \be  \label{resol}
 \sqrt{-\ga} f (\ep) = - {\pi_1 (\ep) + \pi_2 (\ep) \chi  \ov \phi_1 (\ep)+ \phi_2 (\ep) \chi} , \quad
 \sqrt{-\ga} J = {1 \ov \al} {1 \ov \phi_1 + \phi_2 \chi} \ 
 \ee
where $(\chi, \al)$ are determined from $(J_0, f_0)$  by evaluating~\eqref{flowinv} at $z=\ep_0$ 
\be
\begin{split}  
\label{flowint}
\chi &  = - {\phi_1 (\ep_0) \sqrt{-\ga_0} f_0 + \pi_1 (\ep_0) \ov \phi_2 (\ep_0) \sqrt{-\ga_0} f_0 + \pi_2 (\ep_0)},  \cr
\al & = {\phi_2 (\ep_0) \sqrt{-\ga_0} f_0 + \pi_2 (\ep_0) \ov W \sqrt{-\ga_0} J_0} \ .
\end{split}
\ee

Under a change of basis, 
\be
\left( \begin{matrix} \phi_1 \\ \phi_2 \end{matrix} \ri) \to 
\left( \begin{matrix} \tilde \phi_1 \\ \tilde \phi_2 \end{matrix} \ri)=  T \left( \begin{matrix} \phi_1 \\ \phi_2 \end{matrix} \ri), 
\ee
with $T$ a non-singular constant matrix, we have
\be 
\left( \begin{matrix} \al \\ \beta \end{matrix} \ri)  
=  T^t \left( \begin{matrix} \tilde \al \\ \tilde \beta \end{matrix} \ri), \qquad \tilde M = M T^t, \qquad \tilde W = \det T \, W \ .
\ee
It is then manifest from~\eqref{tras} that matrix $\left( \begin{smallmatrix} r&s\\ u&v \end{smallmatrix} \ri)$ is independent of the choice of basis $\phi_{1,2}$.
 Writing $T$ explicitly as 
\be 
T = \left( \begin{matrix} a_+ & b_+ \\ a_- & b_- \end{matrix} \ri)
\ee
we also find that
\be \label{rre1}
\tilde \chi = - {b_+ - a_+ \chi \ov b_- - a_- \chi}, \qquad 
\tilde \al = {\al W \ov \tilde W} (b_- - a_- \chi) \ .
\ee

It seems odd that $\phi_s$ in the above discussion is completely determined by 
boundary conditions (either using the data $f_0$,$J_0$
or equivalently $\chi = \beta/\alpha, \alpha$), whereas one expects
the classical bulk solution $\phi_c$ to not be completely determined 
by boundary conditions at $\epsilon_0$ (we have nowhere imposed boundary
conditions in the IR, which is required to determine the full classical solution.)
The simple resolution to this puzzle is that $\phi_c$ is not the same
as $\phi_s$. In fact one can show using \eqref{bde} and \eqref{pepw}
that $\phi_c$ is determined by $\phi_s$
as follows,

$$
\phi_c(z) \pi_s(z) -\phi_s(z) \pi_c(z) = W(\phi_c, \phi_s) = 1
$$
This determines $\phi_c$ up to the addition of a homogenous solution
$\phi_c \rightarrow \phi_c + c_H \phi_s$.

\subsection{Some examples}

Having laid out the formalism we now turn to a couple of specific examples. We first will examine the flow of the couplings for a free scalar field in AdS$_{d+1}$ and then comment on more general geometries.

\subsubsection{Flow of double-trace couplings in the vacuum} \label{sec:dou}

To gain some intuition for the flow equations let  us first look at the zero momentum sector ($k_\mu =0$) in pure AdS$_{d+1}$ 
\be \label{pnads}
ds^2 = {R^2 \ov z^2} \le(dz^2 + \eta_{\mu \nu} dx^\mu dx^\nu \ri) ,
\ee
for which equation~\eqref{flo} becomes 
\be \label{folw1}
\epsilon\,\p_\ep f = - f^2 - \De \De_- + d\,f
\ee
where we have introduced 
\be
\De = {d \ov 2} + \nu, \qquad \nu = \sqrt{{d^2 \ov 4} + m^2}, \qquad \De_- = d-  \De  \ 
\ee
and will henceforth set $R=1$.  For definiteness, we will consider $\nu \in (0,1)$ so that $\phi$ can be quantized in two ways.\footnote{For $\nu > 1$, the discussion below is still valid. The difference is that $\bar f$ and $\ka_-$ cannot be interpreted as physical couplings any more, but as intermediate steps to obtain the description in standard quantization.} 
In the standard Dirichlet quantization, the corresponding single trace operator dual to $\phi$, which we denote as $\sO_+$, has dimension $\De$, while  in the alternative quantization the corresponding (single trace) boundary operator $\sO_-$, has dimension $\De_-$. Thus in the alternative (standard) quantization the corresponding double-trace coupling has dimension $2 \nu$ ($-2 \nu$).

Writing $f = \bar f + \De_-$, we find that
\be \label{uwo}
\ep \,\p_\ep \bar f = - \bar f^2 + 2 \nu \bar f \ .
\ee
Note that equation~\eqref{uwo} coincides precisely with the double-trace $\beta$-function found in field theories~\cite{Pomoni:2008de, Vecchi:2010jz} for operators 
of dimension $\De_-$. This is consistent with our interpretation of $f$ (up to an additive renormalization $\De_-$) as the double trace coupling in alternative quantization. Similar interpretation was given earlier in~\cite{Vecchi:2010dd}.

We now solve~\eqref{folw1} using the method of sec.~\ref{s:howfs}. 
For pure AdS at $k_\mu =0$, a convenient basis of solutions can be chosen to be
\be
 \phi_1 = z^{\De_-}, \qquad \phi_2 = z^{\De} 
 \ .
 \ee
 Applying~\eqref{resol} and~\eqref{flowint} we  find 
\be
\bar f(\ep) =  {2\nu \ep^{2 \nu} \chi \ov 1 + \chi \ep^{2 \nu}}, \qquad
 J (\ep) = {1 \ov \al}{ \ep^\De \ov 1 + \chi \ep^{2\nu}} \ 
\ee
with
\be \label{intco}
\chi = \ep_0^{-2 \nu} {f_0 - \De_- \ov \De - f_0} , \qquad  
{1 \ov \al} = {2 \nu J_0 \ep_0^{-\De} \ov \De- f_0}  \ .
\ee
$\bar f(\ep)$ has two fixed points (which is clear from~\eqref{uwo}): $\bar f  =0$  in the $\ep \to 0$ limit, which is an UV fixed point, and $\bar f = 2 \nu$ in the limit $\ep \to \infty$, which is the IR fixed point. These two fixed points correspond to the alternative and standard quantizations respectively, as can be seen from the effective conformal dimension of $\bar f$ near each of them (additional additive renormalization for $\bar f$ is required at the IR fixed point).
Below we will refer to them as CFT$^\text{UV}$ and CFT$^\text{IR}$ respectively~(see also Table.~\ref{tabcft}). 

$f_0 = f(\ep_0)$ and $J_0 = J(\ep_0)$ can be considered as bare couplings which depend on the UV cutoff $\ep_0$. From~\eqref{intco}  a continuum limit can be defined as
\be \label{imic}
\bar f (\ep_0) \to  \ka_-  \ep_0^{2 \nu} , \qquad  J (\ep_0) \to J_- \ep_0^\De, \qquad \ep_0 \to 0, 
\ee
with
\be \label{douco}
\ka_- = 2\nu \chi, \qquad J_- = {1 \ov \al} \ 
\ee
interpreted as renormalized (dimensionful) couplings.  As discussed in Appendix~\ref{app:a},  the continuum limit~\eqref{imic}  corresponds to deforming the boundary theory by a double-trace operator given by 
\be \label{alact}
W[\sO_-] = \int \le(J_- \sO_- - \ha \ka_- \sO_-^2 \ri)
\ee
 in alternative quantization. The theory also has an equivalent description in terms of standard quantization with a double trace deformation 
 \be
 W_+ [\sO_+] =  \int \le(J_+ \sO_+ - \ha \ka_+ \sO_+^2 \ri)
 \ee
 with
 \be 
 J_+ =  {J_- \ov \ka_-}, \qquad \ka_+ = -{1 \ov \ka_-} \ 
 \ee 
which is related to~\eqref{alact} by a Legendre transform. 

The above discussion being sufficiently general, of course also applies to near the boundary of AdS$_2$. We have use for this application in Sec.~\ref{s:semiholo}.

\subsubsection{More general geometries: flow in static, rotationally invariant states}

For a general asymptotic AdS metric~\eqref{gmet}
it is convenient to choose the basis of solutions $\phi_{1,2}$ 
that satisfy the asymptotic  behavior 
\be \label{abas}
\phi_1 \to z^{\De_-}, \qquad \phi_2 \to  z^{\De}, \qquad z \to 0 \ .
\ee
We will take the  radial position of the initial surface $\ep_0$, where one defines the field theory, to be small enough so that the geometry there is metrically close to pure AdS~\eqref{pnads}. 
Using~\eqref{resol} we then find the running couplings 
\bea \label{po1}
\sqrt{-\ga}  f(\ep) &=& - {\pi_1 (\ep) + \pi_2 (\ep)  \chi \ov \phi_1 (\ep) + \phi_2 (\ep) \chi}, \cr
\sqrt{-\ga} J (\ep) &=& {1 \ov \al}{1 \ov \phi_1 (\ep) + \phi_2 (\ep) \chi}
\eea
where $\chi$ and $\al$ are again given by~\eqref{intco}. 

Armed with these general solutions one can make the following observations regarding  the flow of the couplings:
\begin{itemize}
\item For $f_0 = \De_-$, i.e., when $\chi=0$, which corresponds to flowing out of the CFT$^\text{UV}$ fixed point with zero bare double-trace couplings, one finds:
\be \label{c1}
\sqrt{-\ga} \, J(\ep) = {J_0 \ov \phi_1 (\ep)} \ep_0^{-\De}, \qquad 
\sqrt{-\ga} \, f(\ep) =- {\pi_1 (\ep) \ov \phi_2 (\ep)} \ .
\ee
\item Likewise, for $f_0 = \De$, i.e. $\chi = \infty$, which corresponds to a flow from  the CFT$^\text{IR}$  fixed point, with zero bare double-trace couplings,
\be \label{c2}
\sqrt{-\ga} \, J(\ep) = {J_0 \ov \phi_2 (\ep)} \ep_0^{-\De_-}, \qquad 
\sqrt{-\ga} \, f(\ep) =- {\pi_2 (\ep) \ov \phi_2 (\ep)}  \ .
\ee
\end{itemize}
For a  generic asymptotic AdS spacetime~\eqref{gmet}, the equations~\eqref{c1}--\eqref{c2}  involve non-trivial functions of $\ep$. This implies that in a generic non-vacuum state (for instance at non-zero temperature or chemical potential), double-trace deformations are generically generated along the renormalization group flow even if one starts with zero bare coupling. The double-trace couplings generated along the flow are precisely the scale-dependent ``response functions'' considered in~\cite{Iqbal:2008by}.

\section{Extremal charged black hole and semi-holography}
 \label{s:semiholo}

\begin{figure}[t]
\begin{center}
\includegraphics[scale = 0.45]{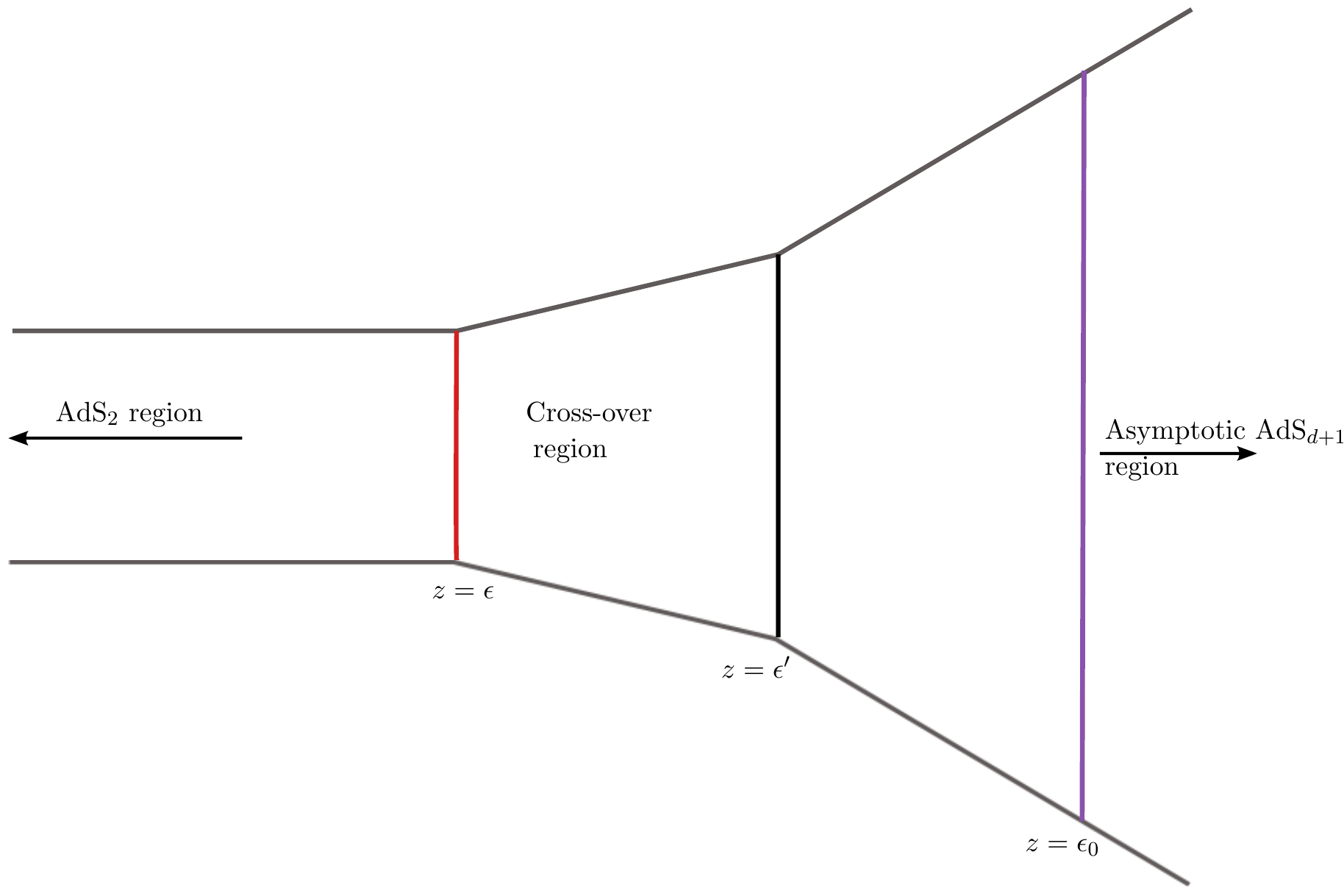}
 \caption{ A simple sketch of extremal black hole geometries with an infinite AdS$_2$ throat and the process of defining effective field theories via the holographic Wilsonian RG flow. To extract low energy physics, it is convenient to integrate out the bulk geometry all the way to 
 the boundary of AdS$_2$, i.e. $z=\ep$ surface in the figure. The induced effective action on 
 $z=\ep$ can be interpreted as multiple-trace deformations of the CFT$_1$ dual to AdS$_2$. }
\label{f:ads2gap}
\end{center}
\end{figure}

We now apply the above discussion to the geometry of an extremal AdS charged black hole, which describes the boundary theory at a finite chemical potential and zero temperature. 
As mentioned in the introduction, the near horizon region of an extremal black hole 
opens up to an AdS$_2$ region. By integrating out the bulk geometry all the way to the boundary of the AdS$_2$, the holographic RG formalism can be used to directly extract low energy behavior from the AdS$_2$ region (see Fig.~\ref{f:ads2gap}). For example it allows a simple derivation of the semi-holographic description for Fermi surface introduced in~\cite{Faulkner:2009wj,Faulkner:2010tq}.  We will use the example of a neutral scalar field for illustration, but the discussion immediately generalizes to charged scalar fields and spinors.

\subsection{Extremal black holes and emergent CFT}

An extremal charged black hole in AdS$_{d+1}$ has a metric given by 
\be
{ds^2 } = \frac{R^2}{z^2}(-h (z)\, dt^2 + d\vec{x}^2) + \frac{R^2}{z^2}\, \frac{dz^2}{h(z)}  \label{RNmetric} \ .
\ee 
The detailed form of function $h$ will not be important for us except that it goes to $1$ as $z \to 0$ (i.e.,  the metric is asymptotically AdS$_{d+1}$) and  has a double zero at the horizon $z_*$ (equivalently a degenerate horizon), with  
 \be
 h (z) \approx \sqrt{d (d-1)}\;{(z_* -z)^2 \ov z_*^2} + \cdots, \qquad z \to z_*  \ .
 \ee 
Thus the black hole geometry interpolates between an AdS$_{d+1}$ with radius $R$ for small $z$ and an AdS$_2 \times \RR^{d-1}$ with AdS$_2$ radius $R_2 = {R \ov \sqrt{d(d-1)}}$ near the horizon. As a result one anticipates that in the IR the system is described by a $(0+1)$-dimensional  CFT$_1$~\cite{Faulkner:2009wj}.  We will call this infrared fixed point, the emergent CFT, or eCFT for short, to distinguish it from the IR fixed point of the double-traced deformed CFT$_d$ introduced in Sec.~\ref{sec:dou}.  

\begin{table}[htdp]
\begin{center}
\begin{tabular}{|c|l|c|}
\hline
{\bf Terminology} & \hspace{0.3cm}{\bf Description} & {\bf Characterization}\\
\hline \hline
CFT$^\text{IR}$ & Boundary CFT$_d$  & $\text{dim}(\sO_+) = \Delta$ \\
&  standard quantization & \\ 
\hline
CFT$^\text{UV}$ & Boundary CFT$_d$ & $\text{dim}(\sO_-) = \Delta_- $\\ 
&  alternative quantization  & \\
\hline
eCFT$^\text{IR}$ & The emergent CFT$_1$  &  $\text{dim}(\Psi_+) = \delta_+ $\\
& standard quantization & \\
\hline 
eCFT$^\text{UV}$ & The emergent CFT$_1$ & 
$\text{dim}(\Psi_-) = \delta_- $\\
& alternative quantization  & \\
\hline
\end{tabular} 
\caption{A taxonomy of various CFTs we encounter in different regions of the spacetime (asymptopia or the AdS$_2$ boundary).}
\label{tabcft}
\end{center}
\end{table}

In the AdS$_2$ region $z \approx z_*$, the equation of motion for $\phi$ becomes  that of
a scalar field in AdS$_2$ with an effective mass square: 
\be
m_k^2 = m^2 + {k^2 \,z_*^2\over R^2}  . \ 
\ee 
Again depending  on the choice of boundary condition (standard or alternative) at the AdS$_2$ boundary, one has the choice of eCFT$^\text{IR}$ or eCFT$^\text{UV}$ in which 
$\phi$ is dual an operator $\Psi_+$ or $\Psi_-$ of conformal dimension
\be 
\de_\pm = \ha \pm \nu_k, \qquad \nu_k = \sqrt{m_k^2 R_2^2 + {1 \ov 4}} \ .
\ee
A short summary of the various conformal fixed points we encounter is provided in Table.~\ref{tabcft}. Note alternative quantization (the $-$ve sign in the above equation) only exists for $\nu_k \in (0,1)$.

We should note that in the AdS$_2$ region it is convenient to consider a different basis of solutions to the scalar equation of motion, viz., take the solutions with near horizon scaling behaviour:
\be \label{altb}
\eta_\pm \to (z_* - z)^{-\de_\mp}, \qquad z \to z_* \ .
\ee
This choice is of course  is related to~\eqref{abas} by a basis transform
\be  \label{bast}
\eta_{\pm} = a_\pm \phi_1 + b_\pm \phi_2 \ ,
\ee
where the coefficients depend on the spatial momenta.

\subsection{Effective action at the boundary of AdS$_2$}

Now let us consider a CFT at  finite chemical potential defined in the UV with a cut-off surface $z=\ep_0$ (which we refer to as the UV surface) in the asymptotic AdS$_{d+1}$ region. 
We may assume that this cut-off theory has double-trace operator turned on.  The double-trace coupling  $\kappa_U$ in CFT$^{\rm UV}$ is related to the bare coupling $f(\ep_0) = f_0$ 
via~\eqref{intco} and~\eqref{douco}, i.e.,
\be \label{pwn}
\ka_{U} = 2 \nu \chi_U, \quad \chi_U =  \ep_0^{-2 \nu}\;  {f_0 - \De_- \ov \De - f_0} \ .
\ee 
We then integrate out the degrees of freedom all the way to a hypersurface (which we  refer to as the IR surface) at $z= \ep \approx z_*$ near the boundary of AdS$_2$ (see Fig.~\ref{f:ads2gap})  Now the effective action $S_B$ on the IR surface can be considered as providing  boundary conditions for fields in the AdS$_2$. From the point of view of the boundary field theory dual to this near horizon geometry $S_B$ corresponds to multiple-trace deformations of the eCFT$_1$. We again consider the free theory, which allows us to restrict attention to  double-trace deformations.  For $\nu_k \in (0,1)$ as discussed in Appendix~\ref{app:a} and Sec.~\ref{sec:dou}, such deformations can be described using two equivalent descriptions, in either the standard or the alternative quantization in the AdS$_2$ region. We will use the description in the alternative quantization below as it is  slightly more convenient. However, when $\nu_k \geq 1$,  the alternative quantization is disallowed, reflecting an important physical difference in this case; we will comment on this at the end once we understand the basic issues.  

Applying ~\eqref{intco} and~\eqref{douco} to AdS$_2$ and its field theory dual, the double-trace couplings in the dual eCFT$^{\rm UV}$ (alternative quantization) can be  expressed in terms of bulk parameters
in $S_B$ defined on the IR surface as\footnote{The factor $\frac{1}{\sqrt{d(d-1)}}$ multiplying $f(\ep)$ in the equation below has its origins in the fact that $f$ is defined in terms of AdS scale $R$ and that the curvature radius between AdS$_{d+1}$ and AdS$_2$ differ by this factor.}
\be
\ka_I (\ep) =  2 \nu_k  \chi_I , \quad \chi_I= {{f(\ep) \ov \sqrt{d (d-1)}} - \de_- \ov \de_+ - {f(\ep) \ov \sqrt{d (d-1)}}}\; (z_*-\ep)^{2 \nu_k} \ .
\ee
Using~\eqref{rre1} and the change of basis~\eqref{bast} one can relate $\chi_I$ to $\chi_U$ as 
\be \label{rror}
\chi_I = -{b_+ - \chi_U \,a_+ \ov b_-- \chi_U \, a_-} \ . 
\ee

\subsection{Gapless modes in the UV region and semi-holography}

Consider for example starting in the UV from the fixed point corresponding to standard quantization of CFT$_d$ ( i.e. with CFT$^\text{IR}_d$) which implies that $\ka_U \to \infty$. We then find using ~\eqref{rror} that the double-trace coupling for the eCFT$^{\rm UV}$ is
\be
\ka_I = -2 \nu_k {a_+ \ov a_-}  \ .
\ee
Suppose that $a_+$ has a zero at some momentum $k_F$.  Then at $k_F$, the system is forced to sit at the unstable UV fixed point eCFT$^{\rm UV}$.  At $k_F$,  $\kappa_I$ has small frequency expansion as 
\be
\ka_I =  c \, \om^2 + \cdots 
\ee
In terms of standard quantization, the corresponding double-trace coupling $\ka_I^{(+)}$ is the inverse of $\ka_I$ (see for instance \eqref{imre}), thus at $k_F$, the effective action in the standard quantization becomes 
\be \label{eirplus}
\ha \int {1 \ov  c\,\om^2 + \cdots}\, \Psi_+^2 
\ee
for the dual operator $\Psi_+$ in eCFT$_1$, which is manifestly non-local. The reason for this non-locality is that,  for $a_+ =0$, in the region between the IR and UV surfaces ($z = \ep$ and $z = \ep_0$ respectively),  there exists a normalizable mode\footnote{Note that for $a_+ =0$, the mode $\eta_+$ is normalizable for both $z \to z_*$ and $z \to 0$.} which corresponds to a gapless mode of the boundary theory CFT$_d$. The non-local nature of~\eqref{eirplus} arises from integrating out such gapless modes. When $\nu_k \in (0,1)$ such non-locality does not cause a problem, as we can describe the same physics using the alternative quantization for AdS$_2$, for which the effective action 
\be \label{eirq}
\ha \int \ka_I \Psi_-^2 
\ee
is perfectly defined at $k=k_F$.  In other words, for $\nu_k \in (0,1)$, there exists an IR description in which the gapless modes in the UV do not play a role. In fact, since we had a-priori assumed to be in the domain where $\nu_k \in (0,1)$, using the language of alternative quantization we arrived at \req{eirq} naturally.

However, for $\nu_k > 1$, we are no longer able to use the alternative quantization in the near horizon AdS$_2$ for eCFT$_1$.  In such a situation one is forced to work with the formalism appropriate for the standard quantization. The above general discussion still applies of course, but  we need to work with \req{eirplus}.
In order to have a local effective action we  should isolate the modes which become gapless in the intermediate region. The way to do this is readily suggested by the bulk action. Writing the bulk action as 
\be 
S = S_0 (z> \ep)+ S_{ct} (z=\ep) + \int_{z=\ep} \ha  \ka_I \,\phi_0^2  
\ee
one now treats the boundary value $\phi_0$ of the bulk field $\phi$ as a source for the boundary theory operator $\Psi_+$.  As discussed in Appendix~\ref{app:a}, $e^{ S_0 + S_{ct}}$ gives the generating functional $\vev{e^{\int \phi_0 \Psi_+} }$ in the standard quantization. Thus at $k=k_F$, the system can be described by the following low energy effective theory
\be \label{rruu}
\ha \int \,  c_2 \, (\p_t \phi_0)^2 + \int \phi_0 \Psi_+
\ee
which is precisely the semi-holographic description.

The physical difference between $\nu_k \in (0,1)$ and $\nu_k \geq 1$ can be understood as follows. The IR contribution to the two point correlation function of $\Psi_+$ is proportional to $\om^{2 \nu_k}$ (this follows from the AdS$_2$ geometry or equivalently from conformal symmetry in eCFT$_1$). When $\nu_k \in (0,1)$ this IR contribution  dominates over the standard analytic contribution $\om^2$ from the UV region. As a result the  gapless mode in the intermediate or UV region does not play a dominant role as $\om \to 0$. This is reflected by the existence of an alternative quantization scheme in which the non-locality does not arise. However, when $\nu_k > 1$, the analytic contribution $\om^2$ dominates over the IR contribution and we can no longer ignore it. Thus we need to explicitly include it in our low energy effective action~\eqref{rruu} rather than integrating it out. 

In the above we have used the example of a neutral scalar for illustrative purposes. One can develop a  parallel story which applies to a charged spinor field for which such a $k_F > 0$ indeed exists for certain range of mass and charge of a spinor field~\cite{Liu:2009dm,Cubrovic:2009ye,Faulkner:2009wj}. The gapless modes in the UV region around $k_F$ can then be interpreted as free fermions (in the large $N$ limit) around a Fermi surface. The main difference for a fermion (or a charged scalar field) is that the window for imposing alternative quantization is $\nu_k \in (0,\ha)$. But the small frequency expansion for these fields now starts at linear in $\om$ rather quadratic. Again we find that the above story applies. Note that in the parameter region $\nu_k \in (0, \ha)$ it is precisely the strong IR contribution which leads to the breakdown of quasi-particle description near the Fermi surface~\cite{Faulkner:2009wj}.

 For  scalar fields, it  can happen that one is able to tune the parameters of the UV CFT, so as to ensure the vanishing of $\kappa_I$ at $k_F =0$. This indicates the onset of an instability and the corresponding gapless modes then describe gapless fluctuations of the order parameter at a quantum critical point.  

More generally, one can fine tune the double-trace coupling (in the appropriate range) $\ka_U$ at the UV surface,  to make the numerator of~\eqref{rror} vanish. Once again this leads to Fermi surfaces for spinors~\cite{Faulkner:2010tq} and quantum phase transitions for scalars as described recently in~\cite{Faulkner:2010gj}.

\section{Vector field and diffusion on the horizon}
\label{s:vectors}

We now turn to the analysis of a vector field in the bulk spacetime, which is dual to a conserved current of the boundary theory. We again derive flow equations for various double-trace couplings. Here the story is more intricate due to presence of gauge modes. In particular, we find that the qualitative features of the effective action $S_B$ depends sensitively on what boundary conditions one imposes at infinity.

For a boundary field theory at a non-zero temperature (which is described by a black hole in the bulk) the low energy behavior of a conserved current is governed by diffusion. As an application for our formalism, we push the cut-off surface all the way to the stretched horizon and show that the diffusion mode can indeed be recovered by coupling the stretched horizon to the effective action $S_B$ coming from integrating out the rest of the geometry. In  achieving this taking into account the gauge symmetries of $S_B$ plays a crucial role. Our derivation may be considered a baby version of a more refined black hole membrane paradigm.

\subsection{Flow equations for vectors}

We consider the following general gauge invariant action in a general background~\eqref{gmet}
\be 
S = S_0 [z>\ep, A_M]
+ S_B [A_M, \ep]  
\ee
where for definiteness we will focus on the Maxwell Lagrangian
\be \label{feadx}
S_0 = -{1 \ov 4}  \int_{z > \ep} d^{d+1} x \sqrt{-g} \   F_{MN} \, F^{MN} \ . 
\ee
As before $S_B$ is the boundary action for the gauge field degrees of freedom living on the cut-off surface $z = \ep$ and again $x^M = (z, x^\mu) = (z, t,x^i)$. 
The equations of motion are just the bulk Maxwell equations
\be \label{maxw}
\p_M \le(\sqrt{-g} \,  F^{MN} \ri) = 0
\ee
with boundary condition 
\be 
\Pi^M \equiv -\sqrt{-g} F^{zM} = {\de S_B \ov \de A_M} \ .
\ee
Setting $M=z$ in the above equation we conclude that
\be 
{\de S_B \ov \de A_z} =0
\ee
for components of the boundary conditions tangential to the boundary, 
\be \label{bdve}
\Pi^\mu \equiv -\sqrt{-g}  F^{z \mu} = {\de S_B \ov \de A_\mu} \ .
\ee
The equations of motion~\eqref{maxw} can be written as a conservation equation
\be \label{cons}
\p_\mu \Pi^\mu = 0
\ee
and an evolution equation
\be \label{eomV}
\p_z \Pi^\mu + \p_\nu \le(\sqrt{-g} F^{\nu \mu} \ri) = 0
\ee

Formally we can proceed as before, with the physical requirement that the  on-shell action and classical solution $A_M$ be independent of $\ep$,  leading to
\bea 
&& 0 = {1 \ov 4}  \int_{z = \ep} d^{d} x \sqrt{-g} \; F_{MN} F^{MN}  
+\p_\ep S_B [{A}_M, \ep]  \cr
&& \qquad+ \int_{z=\ep}\, d^d x\, 
{\de S_B [{A}_\mu] \ov \de {A}_\mu } \, \p_z A_\mu \ 
\label{mmd}
\eea
which shows that the flow is again generated by the Hamiltonian
\be \label{ff2}
\p_\ep  S_B [{ A}_\mu, \ep]  = -\int d^d x\, \sH \ .
\ee
The flow equation can be re-expressed using~\eqref{bdve} as 
\bea
 && \p_\ep S_B[{ A}_\mu,\ep] 
=  -  \int_{z = \ep} d^{d} x \sqrt{-g} \, \le[{1\over 2 \ga}\, g_{\mu\nu} \, {\de S_B \ov \de {A}_\mu } \, 
{\de S_B\ov \de {A}_\nu } \ri. \cr
&&  \qquad \le. + {1 \ov 4}\, F_{\mu \nu} F^{\mu \nu} \ri]  + \int d^d x \, \p_\mu {\de S_B \ov \de {A}_\mu } A_z \ .
\label{vflow} 
\eea
As in the scalar case one should not impose equations of motion for $A_M$ 
in the region $z > \ep$ in evaluating~\eqref{vflow}. 

We now work out some explicit flow equations, expanding $S_B$ again to quadratic level in fields, compatible with the symmetries of the problem.  However, before we proceed to do so, there are two important issues we should bear in mind.

Firstly, an important new element compared with the scalar case is the gauge symmetry.  As a result, we will see below that depending on the specific choice of boundary condition, viz., Dirichlet or Neumann, at the boundary of AdS, there are important differences in the boundary action $S_B$.  

Secondly, note that even though the spacetime has $SO(d)$ rotational symmetry for the geometry~\eqref{gmet}, when we include spatial momentum, the gauge field  breaks this to $SO(d-1)$. This is easy to see since the momentum picks out a direction and the longitudinal and transverse components of the gauge field behave differently and thus should be treated separately. 
Given the spatial momentum we introduce a projector:
\begin{equation}
P_{ij} = \de_{ij} -\frac{k_i \, k_j}{k^2}, \qquad k^2 \equiv \sum_i k_i^2
\label{}
\end{equation}	
using this we can decompose the spatial components of the gauge potential
\bea
&& { A}_i = { A}^T_i + { A}^L {k_i \ov k}  \, \qquad { A}^T_i = P_{ij}\, { A}_j,  \qquad \cr
&&F_{0i}^L = -{i k_i \ov k} (\om {A}^L + k { A}_0)  \equiv - {i k_i \ov k} E^L
\label{ello}
\eea
into its transverse and longitudinal parts.

\subsection{Neumann boundary condition at infinity}

Let us first consider the situation where at infinity we impose Neumann boundary condition
\be \label{neun}
\lim_{z \to 0} \sqrt{-g} F^{z\mu} = J_0^\mu \ 
\ee
as the story is simpler.  We have now taken the UV surface to coincide with the AdS boundary for simplicity, hence $\ep_0 =0$. 
Within the AdS/CFT context these boundary conditions are sensible for $d=3$ and $d=2$~\cite{Witten:2003ya, Marolf:2006nd}.  For $d=3$ (i.e. AdS$_4$) the existence of the Neumann boundary condition can be inferred from the classical electro-magnetic duality of the free Maxwell theory~\cite{Witten:2003ya} and is explored further in~\cite{Leigh:2003ez,Marolf:2006nd,Aharony:2010fk}. 

The boundary condition~\eqref{neun} can be implemented by using the boundary action 
\be 
S_B [A_M, z=0] = \int d^d x \, J^\mu_0 A_\mu 
\ee
and in the bulk path integral, integrating over all values of boundary values $A_\mu (z=0,x)$.
We now integrate out  the gauge field to some hypersurface at $z = \ep$, then the boundary action $S_B$ at $z = \ep$ is obtained by performing the path integral 
\be \label{pah1}
 e^{ i S_B [A_\mu, \ep]} =   \int^{\tilde A_\mu (z=\ep,x) = A_\mu (x)}\!\!\!\!
 [D \tilde A_M] \,  e^{ i S_0 [\tilde A_M] + i \,S_B [ \tilde A_M,z=0]}  .
\ee 
Note that integrating over all boundary values of $\tilde A_M$ at $z=0$ 
promotes $A_\mu$ on the boundary to be a dynamical gauge field (sans kinetic term). 
We will take $J^\mu_0$ to be conserved, i.e., gauge symmetry is preserved also at the boundary $z=0$. 
	
Since the boundary value of $\tilde A_M $ at $z=0$ is not fixed, we can fix the gauge $\tilde A_z=0$ in the path integral~\eqref{pah1} without affecting the boundary value at $z= \ep$. $S_B$ should thus be invariant under the residual gauge symmetry $A_\mu \to A_\mu - \p_\mu \lam (x)$, and as a result satisfies 
\be 
\p_\mu {\de S_B \ov \de {A}_\mu } =0 \ .
\ee
We can then take the boundary action to be given as follows:
\begin{eqnarray}
 S_B [A_\mu, \ep] &=&  \Lambda(\ep) + \int\, {d^dk \over (2\pi)^d} \, \sqrt{-\ga}\;J^\mu(k,\ep) \, A_\mu(-k) \nonumber \\
&&\hspace{-2.27cm}  - \frac{1}{2} \, \int \, {d^dk \over (2\pi)^d} \, \sqrt{-\ga}\, \left[f_T (\ep) \, { A}_T^i  { A}^T_i + h_L(\ep) \, F_{0i}^L F^{0iL} \right] 
\label{sbvec}
\end{eqnarray}
where  $\ga$ was introduced in~\eqref{defge}, $F_{0i}^L$ was introduced  in~\eqref{ello}, and indices in $S_B$ are raised and lowered by induced metric on $z=\ep$ hypersurface.
 Since the spatial part of metric~\eqref{gmet} is proportional to identity matrix, $(A^{T})^{i} = g^{ii} A^T_i$ (no summation) remains transverse, although becoming $\ep$ dependent. 
We require the action~\eqref{sbvec} to be gauge invariant, i.e. $J^\mu$ should satisfy
\be \label{consJ}
\om J^0 = k J^L , \qquad J^L \equiv {J^i k_i \ov k}  \ .
\ee 

First we note that ${ A}^T_i$ is gauge invariant and behaves as a set of decoupled massless scalar fields. Thus we can immediately write down the flow equations for transverse components from~\eqref{lo1}--\eqref{flo},
\bea\label{vJTeq}
&&\hspace{-7mm}{\cal D}_\ep  \le(\sqrt{-\ga} \;{ J}_T^i (k,\ep) \ri)  = -{J}_T^i (k,\ep) \, f_T(k,\ep) 
  \\
&& \hspace{-7mm}{\cal D}_\ep  \le( g^{ii} \sqrt{-\ga}\; f_T(k,\ep)\ri)  = g^{ii} \!\left(- f_T^2 (k,\ep) +  
k_\mu k^\mu \ri) \label{vflowT}
\eea
where ${\cal D}_\ep$ was introduced in~\eqref{defde}. The analysis of these equations is similar to the case of the scalar discussed in Sec.~\ref{s:howfs}. 
The equation for the cosmological constant $\Lam$ is
\be \label{vlameq}
{\cal D}_\ep  \Lam  =  \ha \int {d^d k \ov (2 \pi)^d} \, J^\mu(k,\ep) \,J_\mu(-k,\ep) \ .
\ee

Plugging \eqref{sbvec} into \eqref{vflow} one also finds the following equations for the longitudinal components 
\bea
\label{vJ0eq}
&&\hspace{-3mm}{\cal D}_\ep  \le(\sqrt{-\ga} \;J^0(k, \ep) \ri)  = - k_\mu k^\mu \, h_L (k,\ep)\; J^0 \\
 \label{vflowL}
&&\hspace{-3mm}{\cal D}_\ep  \le( g^{tt}\, g^{ii} \, \sqrt{-\ga}\; h_L (k,\ep)\ri)  = - g^{tt} g^{ii} \left(1  - 
 h_L^2 (k,\ep) \, k_\mu k^\mu \right) \nonumber \\
\eea
where in writing down~\eqref{vJ0eq} we have used~\eqref{consJ}. 

A version of equations~\eqref{vflowT} and~\eqref{vflowL} were derived before~\cite{Iqbal:2008by}, where these equations arose as the flow equations of longitudinal and transverse conductivities. 
This connection also helps to express solutions of~\eqref{vJTeq}--\eqref{vflowL} 
in terms of solutions of classical equations of motion~\eqref{cons}--\eqref{eomV} 
\bea
&& f_T = - g_{ii} {\Pi_T^i \ov \sqrt{-\ga} A_i^T}, \qquad J_T^i = {1 \ov \sqrt{-\ga} A^T_i}
\\
&& h_L = {g_{tt} g_{ii} \ov  \, \omega}  {  \Pi^L  \ov \sqrt{-\ga} \, E^L },  \qquad  
J^0 = {1 \ov \sqrt{-\ga}\, E^L} \ .
\label{}
\eea 
 
Let us look at the lowest order expression for $S_B$ in small $\om$ and $k$ expansion. 
For the transverse flow equation~\eqref{vflowT}, in the limit $\om = k =0$, one should take $f_T =0$. This follows from the fact that one requires $f_T (\ep ) \to 0$ as $\ep \to 0$ (to keep the fixed point theory free of the $A_T^2$ term).  As a result we conclude that $f_T \propto O(k^2, \om^2)$. Now writing 
\be 
\sqrt{-\ga} g^{ii}  f_T =- \lam_0 \,\om^2 +\lam_1 \,k^2 + \cdots
\ee
we find 
\bea \label{0o}
 \lam_0 (\ep) = \lam_0 (\ep_0) + \int_{\ep_0}^\ep  dz \, \sqrt{-g} g^{ii} g^{tt}, \cr
 \lam_1 (\ep) = \lam_1 (\ep_0) +\int_{\ep_0}^\ep  dz \, \sqrt{-g} (g^{ii})^2 
\eea
 Similarly consider~\eqref{vflowL} with $k_\mu =0$, we find for $\ka \equiv g^{tt} g^{ii} \sqrt{-\ga} \, h_L$
\be \label{1o}
\ka (\ep) = \ka (\ep_0) -  \int^\ep_{\ep_0} dz \, \sqrt{-g} g^{tt} g^{ii}  
\ee
We should choose boundary conditions such that the coefficients $\lam_0 (\ep_0),  \lam_1 (\ep_0) , \ka (\ep_0) \to 0$ as $\ep_0 \to 0$. Note that as $z \to 0$, all the integrands in equations~\eqref{0o} and~\eqref{1o} behave as $z^{4-d}$. As a result one obtains  divergent integrals in the limit  $\ep_0 \to 0$ in $d \ge 4$.  This implies that a sensible continuum limit cannot be taken, which appears to be consistent with the conclusions based on normalizability~\cite{Marolf:2006nd}. Note that for $d=2,3$, the inverse of $\lam_{0,1}$ and $\ka$ can be interpreted as the induced gauge coupling. Despite starting with a dynamical gauge theory with no kinetic term in the UV at $z=0$, kinetic terms are generically generated along the flow.

\subsection{Dirichlet boundary condition at infinity}

Let us now consider the situation where at infinity we use the standard Dirichlet boundary condition, i.e., we require
\be
A_\mu (z=0,x) = B_\mu (x)   .
\ee
By the AdS/CFT dictionary, $B_\mu$ is interpreted as an external source coupled to the conserved current $j^\mu$ of the CFT. As such it is important to note that $B_\mu$ is not dynamical and there is no gauge symmetry associated with it. 

We now integrate out $A_M$ to some hypersurface at $z = \ep$, then the boundary action $S_B$ at $z = \ep$ is obtained by performing the path integral
\be \label{pah}
 e^{ i S_B [A_M, \ep]} =   \int^{\tilde A_M (z=\ep,x) = A_M (x)}_{\tilde A_\mu (z=0,x) = B_\mu (x)}
 [D \tilde A_M ]\, e^{ i S_0 [\tilde A_M]} \ .
\ee 
where $S_0$ is given by~\eqref{feadx}.  

We will now set the gauge 
\be
\tilde A_z =0, \qquad z \in [0, \ep]
\ee
by considering a gauge transformation $\lam (z,x)$ which satisfies $\lam (z=0) =0$ (so that $B_\mu$ is unchanged) 
\be 
\lam (z,x) =  \int_{0}^{z} dz' \, \tilde A_z  (z',x) \ . 
\ee
Such a gauge transformation now shifts the upper boundary condition of~\eqref{pah} to
\be \label{ggg}
\tilde A_\mu (z=\ep) = \hat A_\mu \equiv  A_\mu - \p_\mu \varphi , \qquad \varphi(x) = \int_{0}^\ep\,  dz\, \tilde A_z \ .
\ee
The path integral over $\tilde A_z$ now reduces to that over $\vp$ which depends only on $x^\mu$. Thus the left hand side of~\eqref{pah} can be written as 
\be \label{eirp}
\int D \varphi (x) \, e^{S_B [\hat A_\mu, \ep]}  \ .
\ee
$\varphi$ is precisely the ``Goldstone'' mode introduced in~\cite{Nickel:2010pr}.  As emphasized there it is gapless and should be retained in the low energy theory.  This is manifest in~\eqref{eirp} as $\varphi$ appears only with derivatives and we choose not to integrate it out.  Note that the new gauge potential $\hat A_\mu$ introduced above, is gauge invariant, under a residual gauge transformation where one also shifts the value of $\varphi$, i.e.,
\be 
A_\mu \to A_\mu - \p_\mu \lam, \qquad \vp \to \vp - \lam  \ .
\ee
As a consequence $S_B [\hat A_\mu]$ satisfies 
\be 
\p_\mu {\de S_B \ov \de A_\mu} - {\de S_B \ov \de \vp} =0 \ .
\ee 

Our earlier formal derivation of the flow equation~\eqref{vflow} applies
to $S_B [\hat A_\mu]$ inside the path integral for $\vp$; we only need to
replace  $A_\mu$ there  by $\hat A_\mu$.  In contrast to~\eqref{sbvec}, since now $\hat  A_\mu$ is gauge invariant we can introduce 
mass type terms $\hat A_\mu^2$ for the longitudinal and temporal components. 
More explicitly, we can parameterize $S_B[{\hat A}_\mu]$ as (as $\p_\mu \phi$ only shifts longitudinal components of $A_\mu$, we now drop $\hat{}$ for $A_T$ ):
\bwt
\begin{eqnarray}
 S_B [\hat A_\mu, \ep] &=&  \Lambda(\ep) + \int\, {d^dk \over (2\pi)^d} \, \sqrt{-\ga}\; \left(J^\mu(k,\ep) \, {\hat A}_\mu(-k) -\ha\, f_T (k,\ep) \,  A^{\,T}_i (k) A_{\,T}^i (-k) \right) \nonumber \\
&&\quad -  \ha \, \int\, {d^dk \over (2\pi)^d} \, \sqrt{-\ga}\; \left( f_0  (k,\ep) \hat A_0 (k) \hat A^0 (-k) +f_L (k, \ep)  g^{ii} \hat A^L (k)  \hat A^L (-k)   \right) \nonumber \\
&& \quad - \frac{1}{2} \, \int \, {d^dk \over (2\pi)^d} \, \sqrt{-\ga}\, \left[ f_{0L} (\hat A_0 (k) \hat A^L (-k) + \hat A_0 (-k) \hat A^L (k)) \right] 
 \label{sbvec1}
\end{eqnarray} 
\ewt
Since $\hat A_\mu$ is  gauge invariant, $J^\mu$ now does not have to be conserved. 
Plugging~\eqref{sbvec1} into~\eqref{vflow} we find a set of flow equations, which we now proceed to study. 

The flow equations for the transverse components $J_T^i$ and $f_T$ are identical to those in the Neumann boundary case~\eqref{vJTeq}--\eqref{vflowT} as their equations decouple from that for longitudinal and temporal components . For the temporal and longitudinal components we get a set of coupled equations: 
\bea 
&&\sD_\ep (\sqrt{-\ga} J^0) = - J^0 f_0 - g_{ii} J^L f_{0L} \\
&& \sD_\ep (\sqrt{-\ga} J^L) = -  J^L f_L + g_{tt} J^0 f_{0L} 
\eea
and 
\bea 
\label{jevte}
\sD_\ep (\sqrt{-\ga} g^{tt} f_0) &=& - g^{tt} f_0^2 + g_{ii} f_{0L}^2 + g^{tt} g^{ii} k^2  \\
\sD_\ep (\sqrt{-\ga} g^{ii}  f_L) &=& - g^{ii} f_L^2 + g_{tt} f_{0L}^2 - g^{tt} g^{ii} \om^2  \\
\sD_\ep (\sqrt{-\ga} f_{0L}) &=& f_{0L} (f_L + f_{0} ) - g^{tt} g^{ii} \om k
\label{vflow0L}
\eea

Note that if we were to consider a geometry which preserves Lorentz invariance along the boundary directions, i.e., by having $g_{tt} = g_{ii}$ as in the case of pure AdS$_{d+1}$, then $f_0, f_L, f_{0L}$ and $f_T$ collapse into two independent functions whose equations decouple from each other. Here we consider the more general situation as in most example of interest where Lorentz symmetry is broken (finite temperature or chemical potential) . Finally the flow of the cosmological constant $\Lam$ is  as before and is  given by~\eqref{vlameq}.

We again work out the explicit expression for $S_B$ to lowest order in the small $k$ and 
$\om$ expansion. Note that while the flow equations for the transverse components are identical for the Dirichlet and Neumann boundary conditions at infinity, care should be exercised in selecting their solutions. Setting $k=\om=0$ in~\eqref{jevte}--\eqref{vflow0L} and in~\eqref{vflowT} we find the following solution
\bea
\label{sol1}
 &&f_{0L} =0 , \qquad \sqrt{-\ga} \,g^{tt}\, f_0 (\ep) =  {1 \ov Q_t} \\
 &&\sqrt{-\ga} \,g^{ii} \,f_T (\ep) =\sqrt{-\ga} \,g^{ii}\, f_L (\ep) = 
  {1 \ov Q_i} \ 
 \label{sol2}
\eea 
with
\be
Q_t\equiv \int_0^\ep {dz \ov \sqrt{-g} \,g^{zz} \,g^{tt}}, \quad
Q_i = \int_0^\ep {dz \ov \sqrt{-g}\, g^{zz} \,g^{ii}}
\ee
where we have chosen the constants of integration such that $Q_t (\ep), Q_i(\ep) \to 0$ as $\ep \to 0$, the reason for which will be clear momentrily. Recall that in the Neumann case we had to choose $f_T =0$ at this order. One can similarly solve for $J^\mu$ and $\Lam$ by integrating the respective equations. Alternatively,
one can obtain the effective action directly by performing the path integral~\eqref{pah} in the saddle point approximation (which is in fact simpler). One finds that\footnote{The expressions below were obtained earlier~\cite{Nickel:2010pr} and in~\cite{Heemskerk:2010hk} in the pure AdS limit.}
\be
S_B [\hat A_\mu, \ep] = \int 
 \le[{1 \ov 2\, Q_t} \le(\hat A_0 - B_0 \ri)^2 - 
\sum_i {1 \ov 2 \,Q_i} \le(\hat A_i - B_i \ri)^2 \ri]  \label{sbvec3}
\ee
In the cut-off theory defined for $z \geq \ep$, $A_\mu$ is interpreted as the source for the  boundary theory current $j^\mu$. As for the scalar case discussed in Sec.~\ref{s:holwilson} and Appendix~\ref{app:a}, the effective action for the boundary theory is obtained from the Legendre transform of $S_B$ and we find that the corresponding boundary theory 
effective action is 
\be \label{ehje}
I_{UV} = \int 
\le(-\ha Q_t (j^0)^2 + \ha  \sum_i Q_i (j^i)^2 - j^\mu (B_\mu + \p_\mu \varphi) \ri) 
\ee
which has a well defined derivative expansion. Note that integrating out $\varphi$ now imposes the condition that $j^\mu$ is conserved.  One can also now understand our choice of integration constants for the $Q$s in~\eqref{sol1} and~\eqref{sol2} --  these come from the expectation that the double-trace deformations should vanish as we take $\ep$-surface to the boundary since we start at the fixed point without any double-trace deformation.  

If one chooses to integrate out $\varphi$ in~\eqref{sbvec3}, then the resulting $\tilde S_B$ is given by (for simplicity we set $B_\mu =0$ below and $E^L$ was introduced in~\eqref{ello}) 
\bea 
\tilde S_B [ A_\mu, \ep] &=& - \int {d^d k \ov (2 \pi)^d} \, \le[\ha \,{1 \ov Q_i \om^2 -Q_t k^2 } (E^L)^2 \ri. \cr
 && \qquad \qquad\le. + 
\sum_i {1 \ov 2 Q_i} (A_i^T)^2 \ri] \label{sbvec4}
\eea
and is manifestly non-local.  Note that $\tilde S_B$ is gauge invariant and 
satisfies
\be 
\p_\mu {\de \tilde S_B \ov \de A_\mu} =0 \ . 
\ee
Note that~\eqref{sbvec4} has the form of~\eqref{sbvec}. Indeed had we decided to 
integrate out $\vp$ from the beginning, the resulting $\tilde S_B [A_\mu, \ep]$ would be gauge invariant under $A_\mu \to A_\mu - \p_\mu \lam$ and again would be expanded as in~\eqref{sbvec} with coefficients satisfying~\eqref{vJ0eq} and~\eqref{vflowL}.
 
To conclude the discussion let us make a quick comparison of the Dirichlet and Neumann boundary conditions. In the Dirichlet case, apart from the Wilson line mode $\vp$, there are no other gapless modes. This is  natural since we are considering a gauge field on a spacetime 
with an infrared cut-off (and a UV boundary condition), and  the vector spectrum therefore has a  mass gap as in a confining theory. In contrast in the Neumann case, one has a dynamical gauge field which itself is gapless and should be kept in the low energy effective action. This is reflected in the fact that when we do a Legendre transform of~\eqref{sbvec}, we find non-local behavior due to the behavior $f_T \propto O(k^2, \om^2)$ (similarly for the $(F_{0i}^L)^2$ term).  The physical interpretation of these non-local terms is the same as the semi-holographic story described in Sec.~\ref{s:semiholo}.

\subsubsection{Diffusion on the stretched horizon}

As a simple application of the effective action~\eqref{sbvec3} we obtained by integrating out the bulk gauge field in the region $z \in \left(0, \ep\right)$, we derive the diffusion equation for a conserved current at a non-zero temperature. As usual the thermal background is provided by working in a static black hole geometry with a non-degenerate horizon. The discussion follows the one that was recently given in~\cite{Nickel:2010pr}, which we repeat here to highlight the role played by the effective action $S_B$ in this baby version of the refined membrane paradigm.  

  In the absence of external sources, the low energy effective action~\eqref{sbvec3} can be written in momentum space as 
  \be
 S_B [\hat A_\mu, \ep] =  - \ha  \int
  \le(  {1 \ov Q_i}  \hat A_i (k) \hat A_i (-k) -  {1 \ov Q_t}  \hat A_0 (k) \hat A_0 (-k)  \ri)  \label{sbvec2}
\ee
The boundary condition~\eqref{bdve} then becomes 
\be 
\Pi^0 = {\hat A_0 \ov Q_t}, \qquad \Pi^i = - {\hat A_i \ov Q_i}
\ee
The conservation equation for the momentum $\Pi$,~\eqref{cons},  then gives 
\be  \label{op1}
\om {\hat A_0 \ov Q_t} = - k  {\hat A_i \ov Q_i}
\ee
We need to supplement this data with a regularity condition at the horizon; this is essentially the infalling boundary condition and for vectors which relates the conjugate momentum to the electric field a la Ohm's Law~\cite{Damour:1978cg, Thorne:1986iy} (see e.g. Sec.~IIA of~\cite{Iqbal:2008by}
for a review):
\be  \label{op2}
\Pi^i =  \sig F_{ti} = - \sig \, (i k \, \hat A_0 + i \om \, \hat A_i) 
\ee
where $\sig$ is the conductivity. Combining~\eqref{op1} and~\eqref{op2} we then find that 
for
\be \label{eoo}
Q_i \om^2 \ll Q_t k^2 
\ee
 the diffusive dispersion relation: 
\be 
\om = - i D \, k^2, \qquad D = \sig \,Q_t  \ . 
\ee
Note that in the above derivation the use of the effective action is essential. One would not be able to find the diffusion mode using the horizon boundary condition alone. 

For a static black hole geometry, the metric functions behave as: $g_{tt}$,  $g^{zz} \to 0$ with the product $g_{tt} g_{zz}$ remaining constant. Then as $\ep$ approaches the horizon of a black hole, from~\eqref{sol1} and~\eqref{sol2}, $Q_t$ remains finite, while $Q_i$ approaches infinity logarithmically (for a non-degenerate horizon). Thus we cannot put our cut-off surface too close to the horizon
as~\eqref{eoo} will eventually break down. Note that as $Q_i \to \infty$, the coefficients before 
$A_i^2$ terms  in~\eqref{sbvec2} vanish and higher order term in derivative expansion will become important. In such a regime the  the effective action obtained from the Legendre transform of~\eqref{sbvec2} will  be non-local. It would be good to understand the implication of this better.

\section{Conclusions and discussions}
\label{s:conclude}

In this paper we have described a general formalism for developing a holographic Wilsonian renormalization group flow. The basic idea is to mimic the conventional Wilsonian paradigm of quantum field theories in gravity, i.e., starting from a quantum field theory defined with an explicit cut-off, one integrates out momentum shells, so as to define an effective field theory valid for physical processes of interest. Formulating a holographic Wilsonian flow involves integrating out the bulk degrees of freedom between a UV hypersurface on which our field theory is defined and an IR hypersurface which explores the energy scales of interest. A schematic view of this procedure is depicted in Fig.~\ref{f:bbh}.  For illustration we have worked out flow equations for single and double-trace couplings  for scalar operators  and conserved currents in the planar approximation. 

The next important step is to generalize the flow equations to metric, which is 
more complicated. It also requires a better understanding of how to treat gauge degrees of freedom in the bulk. Naively a direct generalization of the scalar and vector story, i.e. equations~\eqref{ff1} and~\eqref{ff2}  to the metric will lead to a flow equation
\be 
\p_\ep S_B [g_{\mu \nu}, \ep] =- \int d^d x \, \sH 
\ee 
with $\sH$ the ADM Hamiltonian density. The ``Goldstone modes'' identified in~\cite{Nickel:2010pr} and the bi-gravity formalisms developed there should also provide important guidance for how to proceed. In particular, as in the vector case discussed earlier we expect that before integrating out these ``Goldstone modes'', the Hamiltonian $\sH$ should not vanish, as emphasized in~\cite{Heemskerk:2010hk}. 
We will leave this for future publication.

One of the applications of the formalism developed here is to provide a convenient 
framework to characterize low energy behavior of a boundary system in terms of 
certain IR region of the bulk spacetime. Examples discussed here include a derivation of the semi-holographic models and diffusion associated with a conserved current at finite temperature (partially following~\cite{Nickel:2010pr}). With the understanding of the flow of the metric one should then be able to  derive the fully fledged hydrodynamic equations on the stretched horizon. In particular, one should also be able to understand the potential extensions to hydrodynamics when extra light degrees are present, as occurs in the studies of extremal black holes~\cite{Edalati:2010hk,Edalati:2010pn} which are dual to zero temperature systems at finite density.

\vspace{0.2in}  
 \centerline{\bf{Acknowledgements}} 
 \vspace{0.2in} 
 
We thank J.~Polchinski and D.~Son for discussions and a preview of their papers. In addition we would like to thank O.~Aharony, R.~Emparan, V.~Hubeny, N.~Iqbal, D.~Marolf, M.~Mezei and S.~Minwalla for related discussions. HL and MR would like to thank the Aspen Centre for Physics for hospitality during the initial stages of this project. MR would also like to thank MIT, Galileo Galilei Institute, Firenze and Imperial College, London for their hospitality during this project.
HL is supported in part by funds provided by the U.S. Department of Energy (D.O.E.) under cooperative research agreement DE-FG0205ER41360 and the OJI program. MR is supported in part by a STFC Rolling Grant and by funds from the INFN. TF is supported by
the Stanford Institute for Theoretical Physics. 

\vspace{0.5cm}

\appendix

\section{Double trace deformation in a CFT and boundary action in AdS} 
\label{app:a}

In this appendix we consider a free scalar field theory in pure AdS to illustrate how to relate the boundary part of the bulk action $S_B$ to the double-trace deformation in the corresponding CFT.  Our discussion is similar in spirit to that of Vecchi in~\cite{Vecchi:2010dd} (see also~\cite{Witten:2001ua,Berkooz:2002ug,Mueck:2002gm,Minces:2002wp,Sever:2002fk}  for earlier work on double-trace deformations). We will work in a geometry of Euclidean signature for simplicity. The scalar action can be written as 
\be \label{scaa}
S_0 = -\ha  \int d^{d+1} x \, \sqrt{g} \, \le( (\p \phi)^2 + m^2 \phi^2 \ri)
\ee
with
\be \label{pads}
ds^2 = {R^2 \ov z^2} \le(dz^2 + \de_{\mu \nu} \,dx^\mu d x^\nu \ri) \ .
\ee
The equation of motion in momentum space takes the form
\be \label{padseom}
z^{d+1} \p_z \le(z^{1-d} \p_z \phi \ri) - k^2 z^2 \phi - m^2 R^2 \phi = 0  \ 
\ee
where now $k^2 = \delta_{\mu\nu}\, k^\mu\,k^\nu$.  We will employ  the following notation:
\bea
&&\De = {d \ov 2} + \nu, \qquad \nu = \sqrt{{d^2 \ov 4} + m^2 R^2} , \nonumber \\
&& \De_- = d - \De = {d \ov 2} - \nu 
\eea
and henceforth work in units where $R=1$.  

For $\nu \in (0,1)$, there are two ways to quantize $\phi$ by imposing Dirichlet or Neumann boundary conditions~\cite{Breitenlohner:1982jf}, which correspond to two different CFTs~\cite{Klebanov:1999tb}. In the standard quantization (Dirichlet boundary condition), the corresponding operator $\sO_+$ has dimension $\De$. In alternative quantization (Neumann boundary condition), the corresponding operator $\sO_-$ has dimension $\De_-$ \cite{Klebanov:1999tb}. We will refer to the corresponding CFTs as  CFT$^\text{IR}_d$ (standard)  and CFT$^\text{UV}_d$ (alternative)  respectively.  The reason for the terminology can be understood as follows: $\sO_-^2$ is a relevant operator in the CFT$^\text{UV}_d$. When we switch this operator on the fixed point corresponding to CFT$^\text{UV}_d$ flows to the the CFT$^\text{IR}_d$ in the IR~\cite{Witten:2001ua}.

\subsection{Standard and alternative quantizations}

Below we first briefly review the procedure for computing correlations functions, for both standard and alternative quantization, to set up our notations, and then extend it to including double-trace deformations. Note that all integrations below which are not explicitly written should be understood in momentum space. 

\paragraph{Standard quantization:} In standard quantization i.e., for CFT$^\text{IR}_d$, the generating functional for the  boundary theory is given by~\cite{Gubser:1998bc,Witten:1998qj} 
\bwt
\be \label{stanD}
e^{I_+ [\phi_0]} \equiv \vev{e^{\int \phi_0 \sO_+} }_+ =  \lim_{\ep \to 0} \int_{\phi_c (\ep) = 
\ep^{\De_-} \phi_0} D \phi \, e^{S_0 [\phi] + S_{ct} [\phi, z=\ep]} = \lim_{\ep \to 0} e^{S_{0} [\phi_c] + S_{ct} [\phi_c]}
\biggr|_{\phi_c (\ep) = 
\ep^{\De_-} \phi_0}
\ee 
\ewt
where $S_0 [\phi_c]$ is the bulk action evaluated on the classical solution $\phi_c$ which is regular at the interior and satisfies the boundary condition 
\be \label{drbd}
\phi_c (\ep) = \ep^{\De_-} \phi_0, \qquad \ep \to 0 \ .
\ee
The `$+$' subscript in $\vev{\cdots}_+$ indicates the expectation value is taken in the  CFT$^\text{IR}$.  We will correspondingly use a `$-$' subscript to indicate correlation functions evaluated in  CFT$^\text{UV}$.
$S_{ct}$ is a counter-term action defined at the cut-off surface $z = \ep$ to make the total action finite. Introduce a basis of solutions $\phi_{1,2} (z; k)$ to~\eqref{padseom} satisfying
\be 
\phi_1 \to z^{\De_-} , \qquad \phi_2 \to z^{\De} \ , \qquad \text{for} \;\;z \to 0
\ee 
with the corresponding canonical momenta along $z$-direction (as defined in~\eqref{bde}) being  $\pi_1, \pi_2$. Then $S_{ct}$ can be found to be (see {\it e.g.},~\cite{Skenderis:2002wp})
\bea 
S_{ct} &=& \ha \int {d^d k \ov (2 \pi)^d}\,  { \pi_1 \ov \phi_1} \, \phi^2 \cr
&=&  - \ha  \int {d^d k \ov (2 \pi)^d}\, \sqrt{\ga} \le(\De_- + {k^2 \ep^2 \ov 2 (1- \nu)} + \cdots \ri)
\phi^2 \nonumber \\
\label{ctac}
\eea
where $\ga$ was introduced in~\eqref{defge} and for~\eqref{pads} $\sqrt{\ga} = z^{-d}$ and the ellipses above denote higher order terms in $k\,\ep$. Denoting a classical solution to~\eqref{padseom} via a linear combination of the basis chosen above, i.e., 
\be \label{repr}
\phi_c = A \, \phi_1 + B \, \phi_2 \ , 
\ee
the boundary condition in the interior of the spacetime determines the ratio
\be \label{aob}
\chi = {B \ov A} \ , 
\ee
while the Dirichlet boundary condition~\eqref{drbd} at $z=\ep$ fixes $A$ to be
\be 
A  ={\ep^{\De_-} \phi_0 \ov \phi_1 (\ep) + \chi \phi_2 (\ep)} = \phi_0 (1 + \cdots)  - \phi_0 \chi \ep^{2 \nu} (1 + \cdots)  . 
\ee
We have refrained from writing out the terms proportional to the momenta above (denoted collectively by the ellipses).
Plugging the above solution to the action and dropping terms which vanish in the $\ep \to 0$ limit, we then find the standard answer for the bulk on-shell action
\be \label{a20}
I_+ [\phi_0] = \lim_{\ep \to 0} \le(S_0 (\phi_c) + S_{ct} (\phi_c, z=\ep) \ri) =
\ha \int {d^d k \ov (2 \pi)^d}\, G_+ \, \phi_0^2 
\ee
with 
\be \label{defgp}
G _+ = 2 \nu \,\chi  \ .
\ee

\paragraph{Alternative quantization:} In the alternative quantization, i.e., for CFT$^\text{UV}$, the generating functional with a source $J_-$ can be found from 
\bwt
\be \label{aldef}
e^{I_-[J_-]} \equiv  \vev{e^{\int J_- \sO_-} }_-=  \lim_{\ep \to 0} \int_{z \geq \ep} D \phi \, e^{S_0 [\phi] + S_{ct} [\phi, z=\ep] + \int_{z=\ep} \sqrt{\ga} \, \tilde J_-  \phi } = \lim_{\ep \to 0} e^{S_{tot} [\phi_c]}
\ee 
\ewt
where on the right hand side we now integrate over all boundary values of $\phi [z=\ep]$. The bulk source $\tilde J_-$ is related to $J_-$ by some powers of $\ep$ which we will determine shortly. In~\eqref{aldef} the same counter-term action $S_{ct}$~\eqref{ctac} will be used and we will see below that it lead to the right boundary condition for alternative quantization.  As usual the last equality in~\eqref{aldef} is obtained by a saddle point approximation with  
\be
S_{tot} [\phi_c] \equiv S_0 [\phi_c] + S_{ct} [\phi_c, z=\ep]  +  \int_{z=\ep} \sqrt{\ga} \; \tilde J_-  \phi_c 
\ee 
where now $\phi_c$ satisfies the Neumann boundary condition (at cut-off surface $z=\ep$)
\be \label{nn1}
\Pi = {\de S_{ct} \ov \de \phi} + \sqrt{\ga} \, \tilde J_-
\ee
$\Pi$ is the canonical momentum for $\phi$ along $z$-direction as introduced in~\eqref{bde}. Writing the general solution $\phi_c$ as in~\eqref{repr}, one has from the Neumann boundary condition \req{nn1} 
\be
A\, \pi_1 + B\, \pi_2 = \frac{\pi_1}{\phi_1} \, (A \phi_1 + B\, \phi_2)  +  \sqrt{\ga} \, \tilde J_- \ . 
\ee
Thus one has the relation\footnote{Note that $\pi_1 \phi_2 - \pi_2 \phi_1 = 2 \nu$, which is the Wronskian of $\phi_1$ and $\phi_2$.}
\bea  
\sqrt{\ga} \, \tilde  J_-
 = -{B \ov \phi_1} (\pi_1 \phi_2 - \pi_2 \phi_1) =- 2 \nu \, {B \ov \phi_1}, \cr
  \quad \to \quad
 B = - {J_- \ov 2 \nu} \le(1 + \cdots \ri)
\label{inalt}
\eea
where we have introduced 
\be 
J_- =  \ep^{-\De} \tilde J_-  \ .
\ee
Equation~\eqref{inalt} gives the usual identification for the alternative quantization. Plugging this solution to $S_{tot}$, we then find that 
\be  \label{algen}
I_-[ J_-] = \lim_{\ep \to 0} S_{tot} [\phi_c]  = \ha \int  \,  G_- \,  J_-^2, 
\ee
with
\be \label{defgm}
 G_- = - {1 \ov 2 \nu \, \chi} \ .
\ee

One can readily check that~\eqref{algen} is related to~\eqref{a20} derived for  the standard quantization by a Legendre transform. This can also be directly seen without any explicit calculation by integrating  both side of~\eqref{stanD} over $\phi_0$ 
\bea \label{alpa}
\int D \phi_0 \, e^{\int \phi_0 J_-} \, \vev{e^{\int \phi_0 \sO_+} }_+ &&\cr
&&  \hspace{-4.5cm}= \lim_{\ep \to 0} \int_{z \geq \ep} D \phi \, e^{S_0 [\phi] + S_{ct} [\phi, z=\ep]+\int_{z=\ep} \sqrt{-\ga} \, \tilde J_-  \phi }
\eea 
with 
\be
J_- =  \ep^{-\De} \tilde J_- , \qquad \phi (z=\ep) = \ep^{\De_-} \phi_0 \ .
\ee 
The right hand side of~\eqref{alpa} is precisely~\eqref{aldef}, while in the saddle point point approximation the left hand side gives the Legendre transform of $I_+ [\phi_0]$ after using~\eqref{stanD}.

On the field theory side, the Legendre transform of a generating functional $I_+ [\phi_0]$ gives the 1PI effective 
potential $\Ga_+ [\Phi_+]$ for the expectation value
\be
\Phi_+ = \vev{\sO_+ \; e^{\int \phi_0 \sO_+ }}_{+} \ .
\ee
We then have the identification
\be \label{ineq}
I_- [J_-] = \Ga_+ [\Phi_+] , \quad {\rm with} \quad J_- \leftrightarrow \Phi_+
\ee 
We conclude that the gravity path integral in~\eqref{aldef} has two  distinct interpretations: (i) as
the  generating functional for  CFT$^\text{UV}$ (alternative quantization) or (ii) as the 1PI effective potential for CFT$^\text{IR}$ (standard quantization). 

Performing a Legendre transform on both sides of~\eqref{ineq}, we also find 
\be \label{newin}
I _+[\phi_0] = \Ga_- [\Phi_-], \quad {\rm with} \quad \phi_0  \leftrightarrow \Phi_- = \vev{\sO_- e^{\int J_- \sO_-}}_{-}
\ee
where $\Ga_- [\Phi_-]$ is the 1PI effective potential for the alternative quantization. Thus the Dirichlet functional integral~\eqref{stanD} also has two interpretations: (i)  as the generating functional for CFT$^\text{IR}$ (standard quantization) or (ii) as  the 1PI effective potential for the CFT$^\text{UV}$ (alternative quantization).

\subsection{Double trace deformations}

Having described the standard story for computing correlation functions in the AdS/CFT context, we now generalize the above discussion by including terms that correspond to deforming the field theory by multi-trace deformations. 

Consider including on both sides of~\eqref{alpa} an additional functional $e^{W [\phi, z= \ep]}$ in the integrand, i.e.,
\bea \label{alpa1}
&&\int D \phi_0 \, e^{\int \phi_0 J_-+ W [\phi_0]} \, \vev{e^{\int \phi_0 \sO_+} }_+ \cr
& =&  \lim_{\ep \to 0} \int_{z \geq \ep} D \phi \, e^{S_0 [\phi] + S_B [\phi, z=\ep]}
\eea 
where on the right hand side we have collected all the boundary terms together into
\be \label{bdact}
S_B [\phi] \equiv S_{ct} [\phi, z=\ep] + W [\phi, z=\ep]+ \int_{z=\ep} \!\!\sqrt{\ga} \, \tilde J_-  \phi  \ .
\ee
For illustration, we consider $W$ quadratic in $\phi$,
\be \label{urrp}
W [\phi] = -\ha \int_{z=\ep}  \, \sqrt{\ga} \, \bar f \, \phi^2 
= -\ha \int \, \ka_- \, \phi_0^2
\ee
with (for $\ep \to 0$)
\be \label{prok}
\ka_- = \bar f\, \ep^{-2 \nu} \ .
\ee
We will show that this corresponds to a double-trace deformation in  the boundary theory~(in both CFT$^\text{UV}$ and CFT$^\text{IR}$).\footnote{Terms of higher powers of $\phi$ in $W[\phi]$  will then correspond to deformations by  higher trace operators.}  

We first consider the left hand side of~\eqref{alpa1}, which we treat in the saddle point approximation.  It has two distinct interpretations:
\begin{enumerate}
\item Performing the $\phi_0$ integral inside the expectation value, we find a saddle point for $\phi_0$ given by
\be 
\phi_0 = {1 \ov \ka_-} (\sO_+ +  J_-)
\ee
and therefore the l.h.s of \req{alpa1} (denoted $\text{l.h.s.}_\eqref{alpa1}$ for brevity), can be written as
\bea
 \text{l.h.s.}_{\eqref{alpa1}} &=& \vev{\exp \le( \int {1 \ov 2 \ka_-} (\sO_+ + J_- )^2 \ri)}_+ \cr
&=&
e^{{J_-^2 \ov 2 \ka_-} } \vev{e^{\int (J_+ \sO_+ - \ha \ka_+ \sO_+^2)}}_+  
\eea
where the source $J_+$ and double-trace coupling $\ka_+$ are given by
\be\label{imre}
J_+ =  {J_- \ov \ka_-}, \qquad \ka_+ = -{1 \ov \ka_-} \ .
\ee
We thus conclude that 
\be 
\log \left( \text{l.h.s.}_{\eqref{alpa1}}\right) = - {J_+^2 \ov 2 \,\ka_+} + I_+ [J_+;\ka_+]
\label{con1}
\ee
where $I_+[J_+;\ka_+]$ is the generating functional (with a source $J_+$) for CFT$^\text{IR}$ with a double-trace deformation with coupling $\ka_+$. 

\item From the discussion around~\eqref{newin}, the l.h.s of~\eqref{alpa1} can also be written as 
\be 
\text{l.h.s.}_\eqref{alpa1} =\int D \phi_0 \, e^{\int \phi_0 J_-} \,e^{\Ga_- [\phi_0] + W [\phi_0] }
\label{new1}
\ee
where $\Ga_- [\phi_0]$ is the 1PI effective potential for the CFT$^{\rm UV}$ with $\phi_0$ interpreted as the expectation value in the presence of some source. Recall in the large $N$ limit if a theory is deformed by an action $W [\sO]$, then the 1PI effective potential simply shifts as  $\Ga [\Phi] \to \Ga [\Phi] + W [\Phi]$ with $\Phi$ the expectation value of $\sO$. Thus we can interpret the exponent of the last factor in~\eqref{new1} as the effective potential $\Ga_-^{(W)} [\phi_0] = \Ga_- [\phi_0] + W [\phi_0]  $ for  CFT$^\text{UV}$ deformed by a double-trace operator 
\be \label{nedo}
W [\sO_-] = -\ha \int \ka_- \sO_-^2  \ . \qquad  
\ee 
Doing the integral over $\phi_0$ by saddle point in~\eqref{new1}, we obtain the Legendre transform of $\Ga_-^{(W)}$, which then gives the generating functional (with a source $J_-$) of the UV CFT deformed by a double-trace action~\eqref{nedo}, i.e. 
\be
\log \left(\text{l.h.s.}_\eqref{alpa1}\right)=  I_- [J_-;\ka_-]
 \label{con2}
\ee
\end{enumerate}

Thus combining the results above, from~\eqref{con1} and~\eqref{con2} we conclude that
\be 
\log \left(\text{l.h.s.}_\eqref{alpa1}\right)  = I_- [J_-;\ka_-] =- {J_+^2 \ov 2\, \ka_+} + I_+ [J_+;\ka_+] \ . 
\label{prkm}
\ee
These quantities are computed on the gravity side by the right hand side of~\eqref{alpa1} to which we now turn.  

In the saddle point approximation the r.h.s.  of~\eqref{alpa1}  (denoted $\text{r.h.s.}_\eqref{alpa1}$) gives 
\be 
\log\left(\text{r.h.s.}_{\eqref{alpa1}}\right) = \lim_{\ep \to 0} 
\le(S_{0} [\phi_c] + S_B [\phi_c, \ep] \ri)
\label{ooo}
\ee
where $\phi_c$ satisfies the Neumann boundary condition 
\be \label{n1}
\Pi = {\de S_{ct} \ov \de \phi} + {\de W \ov \de \phi} + J_- \ep^{-\De_-} \ .
\ee
This implies that for a generic solution~\eqref{repr} and~\eqref{aob}
\be \label{uio}
A \pi_1 + B \pi_2  =\le( {\pi_1 \ov \phi_1} -\sqrt{-\ga}\, \bar f \ri) (A \phi_1 + B \phi_2)  + J_- \ep^{-\De_-}
\ee
from which
\be
A = - {J_- \, \left(1 + {\mathcal O}(k^2 \ep^2)\right) \ov (2 \nu -\bar f ) \chi + {\mathcal O}(k^2 \ep^2) -  \bar f \ep^{-2\nu} (1 + {\mathcal O}(k^2 \ep^2) )}
\ee
A continuum limit can be obtained by taking $\ep \to$ whilst keeping $ \ka_-$ fixed (see~\eqref{prok}), which then yields
\be 
A = -{J_- \ov 2 \nu \,\chi - \kappa_-} \ .
\ee
Now evaluating~\eqref{ooo} we find that 
\be
\log\left(\text{r.h.s.}_{\eqref{alpa1}}\right)  =  \ha \int \, G^{(\ka)}_- \, J_-^2
\ee
with
\be 
G^{(\ka)}_-  = - {1 \ov 2 \nu \chi - \kappa_-} = {1 \ov G^{-1}_- + \kappa_-} \ 
\ee 
where $G_-$ was given in~\eqref{defgm}.  Now equating the left and right hand side
of~\eqref{alpa1} and using~\eqref{prkm} and~\eqref{imre}, we find that
\be \label{ii1}
I_-[J_- ,\ka_-] =   \ha \int \, {1 \ov  G^{-1}_- + \kappa_-} \, J_-^2
\ee
and 
\be \label{ii2}
I_+[J_+ ,\ka_+] =   \ha \int \, {1 \ov G^{-1}_+ + \kappa_+} \, J_+^2
\ee
where $G_+$ was given in~\eqref{defgp}. Note that~\eqref{ii1}--\eqref{ii2} are 
precisely what one expects for double-trace deformation from the field theory side.  See Appendix~\ref{sec:field}  for further details.

\subsection{Summary} \label{ap:su}

To summarize, we have shown that 
\bea \label{finEx}
e^{-{J_+^2 \ov 2 \,\ka_+} + I_+ [J_+;\ka_+]} &=&  e^{I_- [J_-;\ka_-]} \cr
&&  \hspace{-2cm}=\lim_{\ep \to 0} \int_{z \geq \ep} D \phi \, e^{S_0 [\phi] + S_B [\phi, z=\ep]}
\eea 
where $S_0$ is the action~\eqref{scaa} and 
\be
\label{eqsbsimp}
S_B = - \int d^d x \, \sqrt{\ga}  \le[\ha  f (\ep) \,  \phi^2 -  \tilde J_- (\ep) \, \phi \ri],
\ee 
and 
\bea
e^{I_+ [J_+;\ka_+]} \equiv \vev{e^{\int (J_+ \sO_+ - \ha \ka_+ \sO_+^2)}}_+ , \cr
e^{I_- [J_-;\ka_-]} \equiv \vev{e^{\int (J_- \sO_- - \ha \ka_- \sO_-^2)}}_-  \ .
\eea
The relations between various parameters are
\bea
\!\!\!\!\bar f (\ep)  &\equiv& f (\ep) - \De_-, \quad \ka_- = \lim_{\ep \to 0} \bar f\, \ep^{-2 \nu}, \cr
\!\!\!\!J_- &=&   \lim_{\ep \to 0} \ep^{-\De} \tilde J_-,  \;\; J_+ =  {J_- \ov \ka_-}, 
\;\; \ka_+ = -{1 \ov \ka_-}  .
\eea

As $\ka_- \to 0$ (i.e. $\bar f \to 0$), equation~\eqref{finEx} becomes~\eqref{aldef} for alternative quantization and as $\ka_- \to \infty$, the right hand side of~\eqref{finEx}  becomes
\be
\lim_{\ep \to 0} \int_{z \geq \ep} D \phi \, e^{S_0 [\phi] + S_{ct} [\phi, z=\ep]} 
 e^{- {J_+^2 \ov 2 \ka_+}} \delta \le(\phi \ep^{-\De_-} -  J_+ \ri)
\ee
reproducing the standard quantization~\eqref{stanD}.

As discussed in Sec.~\ref{sec:dou} we can use the solution to flow equations to push $S_B (\ep)$ to finite values of $\ep$. Using the expressions there we thus have for any $\ep$ 
\bwt
\be \label{finEx1}
-{J_+^2 \ov 2 \,\ka_+} + I_+ [J_+;\ka_+] = I_- [J_-;\ka_-]  =  \int_{z \geq \ep} D \phi \, e^{S_0 [\phi] + S_B [\phi, z=\ep]} 
\ee
\ewt
where now in \eqref{eqsbsimp} we have,
\be 
\bar f (\ep) = {\ka_- \ep^{2 \nu} \ov 1 + {\ka_- \ov 2 \nu} \ep^{2 \nu}}, \qquad
\tilde J_- (\ep) = {J_- \ep^\De \ov 1 + {\ka_- \ov 2 \nu} \ep^{2\nu}} \ .
\ee

Note that the alternative quantization only applies to $\nu \in (0,1)$. Attempting to quantizing the bulk scalar outside this window one runs into conflict with positivity of energy and unitarity, i.e., CFT$^\text{UV}$ is not an acceptable fixed point  for $\nu  >1$. Equation~\eqref{finEx} still applies to double-trace deformations of the standard quantization, i.e., in CFT$^\text{IR}$.  

The above discussion can be generalized to including higher powers of $\phi$ in $S_B $. We expect that 
\be \label{in1}
\vev{e^{W_+[\sO_+]}}_+ = \vev{e^{W[\sO_-]}}_-=  \lim_{\ep \to 0} \int_{z \geq \ep} D \phi \, e^{S_0 [\phi] + S_B [\phi, z=\ep]}
\ee 
with
\be \label{in2}
W[\sO_-] =\lim_{\ep\to 0} \le(S_B [\phi] - S_{ct} [\phi] \ri) \bigr|_{\phi (\ep) = \ep^{\De_-} \sO_-}
\ee
where counter-term action $S_{ct}$ may also include higher than quadratic powers and $W_+$ is obtained from $W$ by a Legendre transform.

\subsection{Double trace deformation: field theory derivation} \label{sec:field}

Having described the physics of multi trace deformations from the bulk perspective, we now briefly summarize the relevant results for a planar (large $N$) field theory.
Consider a boundary CFT, which we deform by a a double-trace operator
\be \label{defF}
\de S =  - \ha \ka \int d^d x \, \sO^2
\ee
With this deformation the two point function for ${\cal O}$ now becomes:
\bea
G_\ka &=&{1 \ov Z_\ka}  \vev{\sO (x) \sO(0) \, e^{-\ha \ka \int d^d y \, \sO(y)^2}}, \cr
Z_\ka &=& \vev{e^{-\ha \ka \int d^d y \, \sO(y)^2}}
\eea 
which leads to
\bwt
\be
G_\ka (x) = {1 \ov Z_f} \sum_{n=0}^\infty {(-\ka)^n \ov 2^n n!} \le(\prod_{m=1}^n \int d^dy_m\ri)
\vev{\sO(x) \sO(0) \sO(y_1)^2 \cdots \sO (y_n)^2} 
\ee 
\ewt
The disconnected diagrams cancel between up and downstairs, leaving with only connected diagrams. The $n$-th term in the above equation becomes 
\be
(-\ka)^n \int d^dy_1 \cdots d^d y_n \, G(x-y_1) 
\cdots G(y_{n-1}- y_n) G(y_n)
\ee
with $G(x)$ being the Green's function for ${\cal O}$ in the absence of the deformation~\eqref{defF}.
We thus find in momentum space a simple geometric sum
\be
G_\ka (k) = \sum_{n=0}^n (-\ka)^n G^{n+1} (k) = {G (k) \ov 1 + \ka \,G(k)} = {1 \ov G^{-1} (k) +\ka} \ .
\ee


\begin{thebibliography}{10}

\bibitem{Wilson:1973jj}
K.~G. Wilson and J.~B. Kogut, ``{The Renormalization group and the epsilon
  expansion},''
\href{http://dx.doi.org/10.1016/0370-1573(74)90023-4}{{\em Phys. Rept.} {\bf
  12} (1974)  75--200}.

\bibitem{Wegner:1972ih}
F.~J. Wegner and A.~Houghton, ``{Renormalization group equation for critical
  phenomena},'' \href{http://dx.doi.org/10.1103/PhysRevA.8.401}{{\em Phys.Rev.}
  {\bf A8} (1973)  401--412}.

\bibitem{Wilson:1993dy}
K.~G. Wilson, ``{The renormalization group and critical phenomena},''
\href{http://dx.doi.org/10.1103/RevModPhys.55.583}{{\em Rev. Mod. Phys.} {\bf
  55} (1983)  583--600}.

\bibitem{Polchinski:1983gv}
J.~Polchinski, ``{Renormalization and Effective Lagrangians},''
\href{http://dx.doi.org/10.1016/0550-3213(84)90287-6}{{\em Nucl. Phys.} {\bf
  B231} (1984)  269--295}.

\bibitem{Maldacena:1997re}
J.~M. Maldacena, ``{The large N limit of superconformal field theories and
  supergravity},'' \href{http://dx.doi.org/10.1023/A:1026654312961}{{\em Adv.
  Theor. Math. Phys.} {\bf 2} (1998)  231--252},
\href{http://arxiv.org/abs/hep-th/9711200}{{\tt arXiv:hep-th/9711200}}.

\bibitem{Gubser:1998bc}
S.~S. Gubser, I.~R. Klebanov, and A.~M. Polyakov, ``{Gauge theory correlators
  from non-critical string theory},''
  \href{http://dx.doi.org/10.1016/S0370-2693(98)00377-3}{{\em Phys. Lett.} {\bf
  B428} (1998)  105--114},
\href{http://arxiv.org/abs/hep-th/9802109}{{\tt arXiv:hep-th/9802109}}.

\bibitem{Witten:1998qj}
E.~Witten, ``{Anti-de Sitter space and holography},'' {\em Adv. Theor. Math.
  Phys.} {\bf 2} (1998)  253--291,
\href{http://arxiv.org/abs/hep-th/9802150}{{\tt arXiv:hep-th/9802150}}.

\bibitem{Susskind:1998dq}
L.~Susskind and E.~Witten, ``{The holographic bound in anti-de Sitter space},''
\href{http://arxiv.org/abs/hep-th/9805114}{{\tt arXiv:hep-th/9805114}}.

\bibitem{Peet:1998wn}
A.~W. Peet and J.~Polchinski, ``{UV/IR relations in AdS dynamics},''
  \href{http://dx.doi.org/10.1103/PhysRevD.59.065011}{{\em Phys. Rev.} {\bf
  D59} (1999)  065011},
\href{http://arxiv.org/abs/hep-th/9809022}{{\tt arXiv:hep-th/9809022}}.

\bibitem{Akhmedov:1998vf}
E.~T. Akhmedov, ``{A Remark on the AdS / CFT correspondence and the
  renormalization group flow},''
  \href{http://dx.doi.org/10.1016/S0370-2693(98)01270-2}{{\em Phys.Lett.} {\bf
  B442} (1998)  152--158}, \href{http://arxiv.org/abs/hep-th/9806217}{{\tt
  arXiv:hep-th/9806217 [hep-th]}}.

\bibitem{Alvarez:1998wr}
E.~Alvarez and C.~Gomez, ``{Geometric holography, the renormalization group and
  the c theorem},'' \href{http://dx.doi.org/10.1016/S0550-3213(98)00752-4}{{\em
  Nucl.Phys.} {\bf B541} (1999)  441--460},
  \href{http://arxiv.org/abs/hep-th/9807226}{{\tt arXiv:hep-th/9807226
  [hep-th]}}.

\bibitem{Girardello:1998pd}
L.~Girardello, M.~Petrini, M.~Porrati, and A.~Zaffaroni, ``{Novel local CFT and
  exact results on perturbations of N=4 superYang Mills from AdS dynamics},''
  {\em JHEP} {\bf 9812} (1998)  022,
  \href{http://arxiv.org/abs/hep-th/9810126}{{\tt arXiv:hep-th/9810126
  [hep-th]}}.

\bibitem{Distler:1998gb}
J.~Distler and F.~Zamora, ``{Nonsupersymmetric conformal field theories from
  stable anti-de Sitter spaces},'' {\em Adv.Theor.Math.Phys.} {\bf 2} (1999)
  1405--1439, \href{http://arxiv.org/abs/hep-th/9810206}{{\tt
  arXiv:hep-th/9810206 [hep-th]}}.

\bibitem{Balasubramanian:1999jd}
V.~Balasubramanian and P.~Kraus, ``{Space-time and the holographic
  renormalization group},''
  \href{http://dx.doi.org/10.1103/PhysRevLett.83.3605}{{\em Phys.Rev.Lett.}
  {\bf 83} (1999)  3605--3608}, \href{http://arxiv.org/abs/hep-th/9903190}{{\tt
  arXiv:hep-th/9903190 [hep-th]}}.

\bibitem{Freedman:1999gp}
D.~Freedman, S.~Gubser, K.~Pilch, and N.~Warner, ``{Renormalization group flows
  from holography supersymmetry and a c theorem},'' {\em Adv.Theor.Math.Phys.}
  {\bf 3} (1999)  363--417, \href{http://arxiv.org/abs/hep-th/9904017}{{\tt
  arXiv:hep-th/9904017 [hep-th]}}.

\bibitem{deBoer:1999xf}
J.~de~Boer, E.~P. Verlinde, and H.~L. Verlinde, ``{On the holographic
  renormalization group},'' {\em JHEP} {\bf 08} (2000)  003,
\href{http://arxiv.org/abs/hep-th/9912012}{{\tt arXiv:hep-th/9912012}}.

\bibitem{deBoer:2000cz}
J.~de~Boer, ``{The Holographic renormalization group},'' {\em Fortsch.Phys.}
  {\bf 49} (2001)  339--358, \href{http://arxiv.org/abs/hep-th/0101026}{{\tt
  arXiv:hep-th/0101026 [hep-th]}}.


\bibitem{Lewandowski:2002rf}
  A.~Lewandowski, M.~J.~May and R.~Sundrum,
  ``Running with the radius in RS1,''
  Phys.\ Rev.\  D {\bf 67}, 024036 (2003)
  [arXiv:hep-th/0209050].

\bibitem{Lewandowski:2004yr}
  A.~Lewandowski,
  ``The Wilsonian renormalization group in Randall-Sundrum 1,''
  Phys.\ Rev.\  D {\bf 71}, 024006 (2005)
  [arXiv:hep-th/0409192].



\bibitem{sslee}
S.~S.~Lee,
``Holographic Description of Quantum Field Theory,''
Nucl.\ Phys.\  B {\bf 832} (2010) 567
[arXiv:0912.5223 [hep-th]].


\bibitem{Lee:2010ub}
S.~S.~Lee,
``Holographic Description of Large $N$ Gauge Theory,''
arXiv:1011.1474 [hep-th].

\bibitem{Douglas:2010rc}
  M.~R.~Douglas, L.~Mazzucato and S.~S.~Razamat,
  ``Holographic dual of free field theory,''
  arXiv:1011.4926 [hep-th].

\bibitem{Heemskerk:2010hk}
I.~Heemskerk and J.~Polchinski, ``{Holographic and Wilsonian Renormalization
  Groups},''
\href{http://arxiv.org/abs/1010.1264}{{\tt arXiv:1010.1264 [hep-th]}}.

\bibitem{Bredberg:2010ky}
I.~Bredberg, C.~Keeler, V.~Lysov, and A.~Strominger, ``{Wilsonian Approach to
  Fluid/Gravity Duality},''
\href{http://arxiv.org/abs/1006.1902}{{\tt arXiv:1006.1902 [hep-th]}}.

\bibitem{Nickel:2010pr}
D.~Nickel and D.~T. Son, ``{Deconstructing holographic liquids},''
\href{http://arxiv.org/abs/1009.3094}{{\tt arXiv:1009.3094 [hep-th]}}.

\bibitem{Damour:1978cg}
T.~Damour, ``{Black Hole Eddy Currents},''
\href{http://dx.doi.org/10.1103/PhysRevD.18.3598}{{\em Phys. Rev.} {\bf D18}
  (1978)  3598--3604}.

\bibitem{Thorne:1986iy}
K.~Thorne, D.~Macdonald, and R.~Price, {\em {Black holes: The Membrane
  paradigm}}.
\newblock Yale University Press, 1986.

\bibitem{Son:2007vk}
D.~T. Son and A.~O. Starinets, ``{Viscosity, Black Holes, and Quantum Field
  Theory},''
  \href{http://dx.doi.org/10.1146/annurev.nucl.57.090506.123120}{{\em
  Ann.Rev.Nucl.Part.Sci.} {\bf 57} (2007)  95--118},
  \href{http://arxiv.org/abs/arXiv:0704.0240}{{\tt arXiv:arXiv:0704.0240
  [hep-th]}}.

\bibitem{Rangamani:2009xk}
M.~Rangamani, ``{Gravity \& Hydrodynamics: Lectures on the fluid-gravity
  correspondence},''
  \href{http://dx.doi.org/10.1088/0264-9381/26/22/224003}{{\em Class. Quant.
  Grav.} {\bf 26} (2009)  224003},
\href{http://arxiv.org/abs/0905.4352}{{\tt arXiv:0905.4352 [hep-th]}}.

\bibitem{Bhattacharyya:2008jc}
S.~Bhattacharyya, V.~E. Hubeny, S.~Minwalla, and M.~Rangamani, ``{Nonlinear
  Fluid Dynamics from Gravity},''
  \href{http://dx.doi.org/10.1088/1126-6708/2008/02/045}{{\em JHEP} {\bf 02}
  (2008)  045},
\href{http://arxiv.org/abs/0712.2456}{{\tt arXiv:0712.2456 [hep-th]}}.

\bibitem{Iqbal:2008by}
N.~Iqbal and H.~Liu, ``{Universality of the hydrodynamic limit in AdS/CFT and
  the membrane paradigm},''
  \href{http://dx.doi.org/10.1103/PhysRevD.79.025023}{{\em Phys. Rev.} {\bf
  D79} (2009)  025023},
\href{http://arxiv.org/abs/0809.3808}{{\tt arXiv:0809.3808 [hep-th]}}.

\bibitem{Faulkner:2009wj}
T.~Faulkner, H.~Liu, J.~McGreevy, and D.~Vegh, ``{Emergent quantum criticality,
  Fermi surfaces, and AdS2},''
\href{http://arxiv.org/abs/0907.2694}{{\tt arXiv:0907.2694 [hep-th]}}.

\bibitem{Liu:2009dm}
H.~Liu, J.~McGreevy, and D.~Vegh, ``{Non-Fermi liquids from holography},''
\href{http://arxiv.org/abs/0903.2477}{{\tt arXiv:0903.2477 [hep-th]}}.

\bibitem{Cubrovic:2009ye}
M.~Cubrovic, J.~Zaanen, and K.~Schalm, ``{String Theory, Quantum Phase
  Transitions and the Emergent Fermi-Liquid},''
  \href{http://dx.doi.org/10.1126/science.1174962}{{\em Science} {\bf 325}
  (2009)  439--444},
\href{http://arxiv.org/abs/0904.1993}{{\tt arXiv:0904.1993 [hep-th]}}.

\bibitem{Faulkner:2010da}
T.~Faulkner, N.~Iqbal, H.~Liu, J.~McGreevy, and D.~Vegh, ``{From black holes to
  strange metals},''
\href{http://arxiv.org/abs/1003.1728}{{\tt arXiv:1003.1728 [hep-th]}}.

\bibitem{Faulkner:2010tq}
T.~Faulkner and J.~Polchinski, ``{Semi-Holographic Fermi Liquids},''
\href{http://arxiv.org/abs/1001.5049}{{\tt arXiv:1001.5049 [hep-th]}}.

\bibitem{Gubser:2009cg}
S.~S. Gubser and A.~Nellore, ``{Ground states of holographic
  superconductors},'' \href{http://dx.doi.org/10.1103/PhysRevD.80.105007}{{\em
  Phys.Rev.} {\bf D80} (2009)  105007},
  \href{http://arxiv.org/abs/arXiv:0908.1972}{{\tt arXiv:arXiv:0908.1972
  [hep-th]}}.

\bibitem{Horowitz:2009ij}
G.~T. Horowitz and M.~M. Roberts, ``{Zero Temperature Limit of Holographic
  Superconductors},''
  \href{http://dx.doi.org/10.1088/1126-6708/2009/11/015}{{\em JHEP} {\bf 0911}
  (2009)  015}, \href{http://arxiv.org/abs/arXiv:0908.3677}{{\tt
  arXiv:arXiv:0908.3677 [hep-th]}}.

\bibitem{Gauntlett:2009bh}
J.~P. Gauntlett, J.~Sonner, and T.~Wiseman, ``{Quantum Criticality and
  Holographic Superconductors in M-theory},''
  \href{http://dx.doi.org/10.1007/JHEP02(2010)060}{{\em JHEP} {\bf 1002} (2010)
   060}, \href{http://arxiv.org/abs/arXiv:0912.0512}{{\tt arXiv:arXiv:0912.0512
  [hep-th]}}.

\bibitem{Li:2000ec}
M.~Li, ``{A note on relation between holographic RG equation and Polchinski's
  RG equation},'' \href{http://dx.doi.org/10.1016/S0550-3213(00)00201-7}{{\em
  Nucl. Phys.} {\bf B579} (2000)  525--532},
\href{http://arxiv.org/abs/hep-th/0001193}{{\tt arXiv:hep-th/0001193}}.

\bibitem{Akhmedov:2002gq}
E.~T. Akhmedov, ``{Notes on multi-trace operators and holographic
  renormalization group},''
\href{http://arxiv.org/abs/hep-th/0202055}{{\tt arXiv:hep-th/0202055}}.

\bibitem{Pomoni:2008de}
E.~Pomoni and L.~Rastelli, ``{Large N Field Theory and AdS Tachyons},''
  \href{http://dx.doi.org/10.1088/1126-6708/2009/04/020}{{\em JHEP} {\bf 04}
  (2009)  020},
\href{http://arxiv.org/abs/0805.2261}{{\tt arXiv:0805.2261 [hep-th]}}.

\bibitem{Akhmedov:2010sw}
E.~Akhmedov and E.~Musaev, ``{An exact result for Wilsonian and Holographic
  renormalization group},''
  \href{http://dx.doi.org/10.1103/PhysRevD.81.085010}{{\em Phys.Rev.} {\bf D81}
  (2010)  085010}, \href{http://arxiv.org/abs/1001.4067}{{\tt arXiv:1001.4067
  [hep-th]}}.

\bibitem{Vecchi:2010jz}
L.~Vecchi, ``{The Conformal Window of deformed CFT's in the planar limit},''
  \href{http://dx.doi.org/10.1103/PhysRevD.82.045013}{{\em Phys. Rev.} {\bf
  D82} (2010)  045013},
\href{http://arxiv.org/abs/1004.2063}{{\tt arXiv:1004.2063 [hep-th]}}.

\bibitem{Breitenlohner:1982jf}
P.~Breitenlohner and D.~Z. Freedman, ``{Stability in Gauged Extended
  Supergravity},''
\href{http://dx.doi.org/10.1016/0003-4916(82)90116-6}{{\em Ann. Phys.} {\bf
  144} (1982)  249}.

\bibitem{Klebanov:1999tb}
I.~R. Klebanov and E.~Witten, ``{AdS/CFT correspondence and symmetry
  breaking},'' \href{http://dx.doi.org/10.1016/S0550-3213(99)00387-9}{{\em
  Nucl. Phys.} {\bf B556} (1999)  89--114},
\href{http://arxiv.org/abs/hep-th/9905104}{{\tt arXiv:hep-th/9905104}}.

\bibitem{Iqbal:2010eh}
N.~Iqbal, H.~Liu, M.~Mezei, and Q.~Si, ``{Quantum phase transitions in
  holographic models of magnetism and superconductors},''
  \href{http://dx.doi.org/10.1103/PhysRevD.82.045002}{{\em Phys. Rev.} {\bf
  D82} (2010)  045002},
\href{http://arxiv.org/abs/1003.0010}{{\tt arXiv:1003.0010 [hep-th]}}.

\bibitem{Faulkner:2010gj}
T.~Faulkner, G.~T. Horowitz, and M.~M. Roberts, ``{Holographic quantum
  criticality from multi-trace deformations},''
\href{http://arxiv.org/abs/1008.1581}{{\tt arXiv:1008.1581 [hep-th]}}.

\bibitem{Papadimitriou:2010as}
I.~Papadimitriou, ``{Holographic renormalization as a canonical
  transformation},''
\href{http://arxiv.org/abs/1007.4592}{{\tt arXiv:1007.4592 [hep-th]}}.

\bibitem{Vecchi:2010dd}
L.~Vecchi, ``{Multitrace deformations, Gamow states, and Stability of
  AdS/CFT},''
\href{http://arxiv.org/abs/1005.4921}{{\tt arXiv:1005.4921 [hep-th]}}.

\bibitem{Witten:2003ya}
E.~Witten, ``{SL(2,Z) action on three-dimensional conformal field theories with
  Abelian symmetry},'' \href{http://arxiv.org/abs/hep-th/0307041}{{\tt
  arXiv:hep-th/0307041 [hep-th]}}.

\bibitem{Marolf:2006nd}
D.~Marolf and S.~F. Ross, ``{Boundary Conditions and New Dualities: Vector
  Fields in AdS/CFT},'' {\em JHEP} {\bf 0611} (2006)  085,
  \href{http://arxiv.org/abs/hep-th/0606113}{{\tt arXiv:hep-th/0606113
  [hep-th]}}.

\bibitem{Leigh:2003ez}
R.~G. Leigh and A.~C. Petkou, ``{SL(2,Z) action on three-dimensional CFTs and
  holography},'' {\em JHEP} {\bf 12} (2003)  020,
\href{http://arxiv.org/abs/hep-th/0309177}{{\tt arXiv:hep-th/0309177}}.

\bibitem{Aharony:2010fk}
O.~Aharony, D.~Marolf, and M.~Rangamani, ``{to appear},''.

\bibitem{Edalati:2010hk}
M.~Edalati, J.~I. Jottar, and R.~G. Leigh, ``{Shear Modes, Criticality and
  Extremal Black Holes},''
  \href{http://dx.doi.org/10.1007/JHEP04(2010)075}{{\em JHEP} {\bf 1004} (2010)
   075}, \href{http://arxiv.org/abs/arXiv:1001.0779}{{\tt arXiv:arXiv:1001.0779
  [hep-th]}}.

\bibitem{Edalati:2010pn}
M.~Edalati, J.~I. Jottar, and R.~G. Leigh, ``{Holography and the sound of
  criticality},'' \href{http://arxiv.org/abs/arXiv:1005.4075}{{\tt
  arXiv:arXiv:1005.4075 [hep-th]}}.

\bibitem{Witten:2001ua}
E.~Witten, ``{Multi-trace operators, boundary conditions, and AdS/CFT
  correspondence},''
\href{http://arxiv.org/abs/hep-th/0112258}{{\tt arXiv:hep-th/0112258}}.

\bibitem{Berkooz:2002ug}
M.~Berkooz, A.~Sever, and A.~Shomer, ``{'Double trace' deformations, boundary
  conditions and space-time singularities},'' {\em JHEP} {\bf 0205} (2002)
  034, \href{http://arxiv.org/abs/hep-th/0112264}{{\tt arXiv:hep-th/0112264
  [hep-th]}}.

\bibitem{Mueck:2002gm}
W.~Mueck, ``{An Improved correspondence formula for AdS / CFT with multitrace
  operators},'' \href{http://dx.doi.org/10.1016/S0370-2693(02)01487-9}{{\em
  Phys.Lett.} {\bf B531} (2002)  301--304},
  \href{http://arxiv.org/abs/hep-th/0201100}{{\tt arXiv:hep-th/0201100
  [hep-th]}}.

\bibitem{Minces:2002wp}
P.~Minces, ``{Multitrace operators and the generalized AdS / CFT
  prescription},'' \href{http://dx.doi.org/10.1103/PhysRevD.68.024027}{{\em
  Phys.Rev.} {\bf D68} (2003)  024027},
  \href{http://arxiv.org/abs/hep-th/0201172}{{\tt arXiv:hep-th/0201172
  [hep-th]}}.

\bibitem{Sever:2002fk}
A.~Sever and A.~Shomer, ``{A Note on multitrace deformations and AdS/CFT},''
  {\em JHEP} {\bf 0207} (2002)  027,
  \href{http://arxiv.org/abs/hep-th/0203168}{{\tt arXiv:hep-th/0203168
  [hep-th]}}.

\bibitem{Skenderis:2002wp}
K.~Skenderis, ``{Lecture notes on holographic renormalization},''
  \href{http://dx.doi.org/10.1088/0264-9381/19/22/306}{{\em Class. Quant.
  Grav.} {\bf 19} (2002)  5849--5876},
\href{http://arxiv.org/abs/hep-th/0209067}{{\tt arXiv:hep-th/0209067}}.

\end{thebibliography}

\providecommand{\href}[2]{#2}\begingroup\raggedright\endgroup

\end{document}